\DocumentMetadata{
 lang = en-US,
 pdfversion = 2.0,
 pdfstandard = ua-2,
 tagging = on,
 tagging-setup = {math/setup=mathml-SE}
 }
 
\documentclass[acmsmall,screen]{acmart}

\AtBeginDocument{%
  \providecommand\BibTeX{{%
    Bib\TeX}}}

\setcopyright{acmlicensed}
\copyrightyear{2018}
\acmYear{2018}
\acmDOI{XXXXXXX.XXXXXXX}

\acmJournal{JACM}
\acmVolume{37}
\acmNumber{4}
\acmArticle{111}
\acmMonth{8}

\usepackage{fancyvrb}
\usepackage{tabularx}
\usepackage{float}
\usepackage{xcolor}
\usepackage{booktabs}
\usepackage{siunitx}
\usepackage{lscape}
\newif\ifjustaccepted
\justacceptedtrue   
\ifjustaccepted
  \newcommand\JustAcceptedMark{%
    \begin{tikzpicture}
      \node[rotate=45, scale=4, text=gray, text opacity=0.30,
            font=\normalfont\Huge\bfseries] {Just Accepted};
    \end{tikzpicture}}
  \ExplSyntaxOn
  \AddToHook{shipout/foreground}
    {
      \tag_mc_begin:n {artifact}   
      \tag_stop:                   
      \put(0.5\paperwidth,-0.5\paperheight){\makebox(0,0){\JustAcceptedMark}}
      \tag_start:
      \tag_mc_end:
    }
  \ExplSyntaxOff
\fi

\definecolor{codekeyword}{HTML}{194A8D} 
\definecolor{codestring}{HTML}{71398A}  
\definecolor{codecomment}{HTML}{21643A} 
\definecolor{codefunction}{HTML}{194A8D} 

\usepackage{tcolorbox}

\usepackage{iftex}
\ifLuaTeX
  \usepackage{fontspec}
  \newfontfamily\FAsolid{FontAwesome5Free-Solid-900.otf}
  \newfontfamily\FAregular{FontAwesome5Free-Regular-400.otf}
  \ExplSyntaxOn
  \NewDocumentCommand\decor{m}
    { \mode_leave_vertical: \tag_mc_artifact_group_begin:n {notype} #1 \tag_mc_artifact_group_end: }
  \NewDocumentCommand\iconalt{mm}
    { \mode_leave_vertical: \tag_struct_begin:n {tag=Span, actualtext={#2}}
      \tag_mc_begin:n {} #1 \tag_mc_end: \tag_struct_end: }
  \ExplSyntaxOff
  \newcommand\faUser{\decor{{\FAsolid\symbol{"F007}}}}
  \newcommand\faQuestionCircle{\decor{{\FAsolid\symbol{"F059}}}}
  \newcommand\faLightbulbO{\decor{{\FAregular\symbol{"F0EB}}}}
\else
  \newcommand\decor[1]{#1}
  \newcommand\iconalt[2]{#1}
  \newcommand\faUser{}
  \newcommand\faQuestionCircle{}
  \newcommand\faLightbulbO{}
\fi
\tcbuselibrary{skins,breakable}
\definecolor{boxaccent}{HTML}{264F73} 
\definecolor{boxbackground}{HTML}{F3F6F9}
\definecolor{boxoutline}{HTML}{8B98A4}
\tcbset{
  rqstyle/.style={
    enhanced,breakable,
    frame hidden,boxrule=0pt,arc=0pt,
    colback=boxbackground,
    borderline west={2pt}{0pt}{boxaccent},
    boxsep=0pt,left=10pt,right=9pt,top=7pt,bottom=7pt,
    before skip=0.6\baselineskip,after skip=0.6\baselineskip,
    fontupper=\itshape
  },
  summarystyle/.style={
    enhanced,breakable,
    colback=white,colframe=boxoutline,boxrule=0.7pt,arc=2pt,
    boxsep=0pt,left=9pt,right=9pt,top=7pt,bottom=7pt,
    before skip=0.6\baselineskip,after skip=0.6\baselineskip,
    fontupper=\small
  },
  promptstyle/.style={
    enhanced,breakable,
    colback=boxbackground,colframe=boxaccent,
    colbacktitle=boxaccent,coltitle=white,
    boxrule=0.65pt,arc=2pt,
    boxsep=0pt,left=9pt,right=9pt,top=8pt,bottom=7pt,
    before skip=0.7\baselineskip,after skip=0.7\baselineskip,
    fonttitle=\bfseries\small,fontupper=\small
  }
}
\newtcolorbox{framed}{rqstyle}
\newtcolorbox{custombox}{summarystyle}
\newcounter{promptbox}[section]

\NewDocumentEnvironment{llmprompt}{O{} m}{%
  \refstepcounter{promptbox}%
  \begin{tcolorbox}[promptstyle,title={#1}]%
  \label{#2}%
}{%
  \end{tcolorbox}%
}

\def\BibTeX{{\rm B\kern-.05em{\sc i\kern-.025em b}\kern-.08emT\kern-.1667em\lower.7ex\hbox{E}\kern-.125emX}}

\newcounter{accessiblecode}
\renewcommand{\theaccessiblecode}{\arabic{accessiblecode}}
\newcommand{\accessiblecodecaption}[2]{%
  \refstepcounter{accessiblecode}%
  \noindent\textbf{Listing~\theaccessiblecode: #1}\label{#2}\par\smallskip
}

\begin{document}

\title[Program Readability and Comprehension]{An Empirical Study on the Impact of String Interpolation on Program Readability and Comprehension}


\author{Daniel Ogenrwot}
\email{ogenrwot@unlv.nevada.edu}
\affiliation{%
    \institution{University of Nevada Las Vegas}
    \city{Las Vegas}
    \state{Nevada}
    \country{USA}
    }
\orcid{0000-0002-0133-8164}

\author{John Businge}
\email{john.businge@unlv.edu}
\affiliation{
    \institution{University of Nevada Las Vegas}
    \city{Las Vegas}
    \state{Nevada}
    \country{USA}
    }
\orcid{0000-0003-3206-7085}

\author{Andreas Stefik}
\email{andreas.stefik@unlv.edu}
\affiliation{
    \institution{University of Nevada Las Vegas}
    \city{Las Vegas}
    \state{Nevada}
    \country{USA}
    }
\orcid{0000-0002-2554-5862}

\renewcommand{\shortauthors}{D. Ogenrwot et al.}

\newcommand{\AbstractCategory}[1]{%
  \par\addvspace{.5\baselineskip}
  \noindent\textbf{#1}\quad\ignorespaces
}

%
\begin{abstract}

\AbstractCategory{Objectives.}
Literal string interpolation is a widely used programming language construct that allows developers to embed variables, expressions, and functions directly within strings (e.g., \texttt{f``Hello, \{name\}!''}), reducing the need for explicit concatenation. Compared to traditional concatenation, it is often considered more concise and expressive. However, its effect on program readability and comprehension remains underexplored. The objective of this study is to empirically examine how literal string interpolation influences program readability and comprehension relative to conventional string concatenation.

\AbstractCategory{Participants.}
A total of 314 participants took part in the study, representing a broad range of educational and professional stages. All participants possessed prior experience with at least one programming language, ensuring baseline familiarity with programming syntax and logic structures relevant to the study tasks. While the majority were recruited through our Department of Computer Science, additional participants were reached via a public broadcast shared on LinkedIn, which expanded recruitment beyond the immediate university community.

\AbstractCategory{Study Methods.}
We conducted a randomized controlled experiment using a within-subjects repeated-measures design. Data were collected through a custom-built experimental tool designed to present randomized programming tasks under two conditions: concatenation and interpolation. For each task, two dependent variables i.e. completion time and response correctness, were automatically recorded to measure performance and comprehension accuracy.

\AbstractCategory{Findings.}
String concatenation, particularly in complex expressions, significantly reduced program readability and increased comprehension time. A repeated-measures ANOVA revealed a significant main effect of \textit{task category}, $F(1, 5026) = 92.35$, $p < .001$, $\eta^2_G = .018$, indicating differences in completion time between concatenation and interpolation tasks. There was also a smaller but statistically significant main effect of \textit{academic level}, $F(6, 5026) = 2.41$, $p = .025$, $\eta^2_G = .003$. The interaction between task category and academic level was not significant, $F(6, 5026) = 0.03$, $p = 1.00$, $\eta^2_G = .00$, suggesting that the effect of task category was consistent across levels of academic experience. Accuracy was generally high, with a moderate positive correlation between participant and correct responses (\(r = .55, p < .001\)), indicating consistent task understanding. Additionally, qualitative findings showed that participants viewed interpolation as cleaner and more intuitive, whereas concatenation was preferred mainly due to familiarity. Readability perceptions were shaped by experience and context, with participants emphasizing the importance of clear formatting, variable highlighting, and IDE support, indicating that readability depends as much on tooling and conventions as on syntax.

\AbstractCategory{Conclusions.}
The findings suggest that literal string interpolation can enhance code readability and reduce cognitive effort during comprehension tasks. These insights have implications for programming language design and may inform future standardization of string interpolation syntax to improve code clarity and developer productivity.

\end{abstract}



\setcopyright{cc}
\setcctype{by}
\acmJournal{TOCE}
\acmYear{2026} \acmVolume{1} \acmNumber{1} \acmArticle{}
\acmMonth{1} \acmDOI{10.1145/3872269}
\received{31 December 2025}
\received[revised]{10 August 2026}
\received[accepted]{5 November 2026}

%
\maketitle



\section{Introduction}
\subsection{Problem}
In computer science education, students often learn to use, reason about, and sometimes invent programming languages~\cite{10.1145/3159450.3159488,Chen02012019,10.1145/2534973,10.1145/3531056.3542775}. In industry, while programming is hardly the only activity, it is a common and arguably important one. Given that programming languages are large and complex systems, it is no surprise that their impact on people is also complex. For example, studies show that syntax has many complex impacts~\cite{10.1145/2534973} and those impacts are not always universal~\cite{10.1145/3344429.3372508}. Further, as students learn and age, the impact of competing designs can influence them (e.g., with lambdas~\cite{10.1145/2884781.2884849}, compiler errors).

One area students commonly learn right away, especially at the college level, is strings and string processing. Indeed, the first program many students write is often some version of ``Hello, World!'' While students often have to write strings, exactly how they work varies substantially across programming languages. In this study, we investigate one aspect of these differences.

Beyond a first program, the string data type is a crucial element in programming languages~\cite{hojjat2019strings,Kao2025,10.1145/2687148.2687152}, widely recognized for its versatility and essential role in various programming scenarios and paradigms. Strings are fundamental for handling a multitude of text-based operations. In the course of developing software, programmers often engage with strings extensively, adjusting and combining them to achieve a desired output. One approach to modification is inserting variables or expressions within a string literal---sometimes called String Interpolation~\cite{pep_python}. \citet{10.1145/3173574.3174101} proposed the use of string interpolation as a solution to hard to refactor code snippets.

While string interpolation offers a concise alternative to concatenation, whether it actually benefits students or those with more experience is arguable. For example, it might introduce code complexity, particularly when handling complex expressions or multi-line strings. This complexity might adversely affect code readability, making it more challenging to comprehend, debug, or maintain. Similar to string concatenation, string interpolation can also introduce security vulnerabilities. When user-provided data is improperly escaped or sanitized, it can lead to several security vulnerabilities, including XML External Entity (XXE) injection, SQL injection, Cross-Site Scripting (XSS) and script injection attacks~\cite{10.1145/2786805.2786821, 10.1145/2687148.2687152}. In addition, programming languages have not standardized common rules around string manipulation. For instance, Python utilizes the \textit{f-string} format, which differs from the syntax used in Java, Rust, Scala, C, or C++. These lexical differences, including how string literals are delimited, also complicate tools that analyze code across languages~\cite{ogenrwot2026pyreprism}. Listings~\ref{lst:java}--\ref{lst:scala} then show how these languages can produce the same output with different syntax, some through placeholders embedded in the string and others through operator-based composition. To make the central contrast concrete, Figure~\ref{fig:side-by-side} shows the same output built by concatenation and by interpolation. 

\noindent\begin{minipage}[t]{0.48\linewidth}
\accessiblecodecaption{Java}{lst:java}
\begin{Verbatim}[commandchars=|\[\],fontsize=\footnotesize]
|codecm[/* code snippet */]
|codekw[String] name = |codestr["World"];
|codekw[String] greeting =
  |codekw[String].|codefunc[format](|codestr["Hello, %s!"], name);
|codefunc[System.out.println](greeting);
\end{Verbatim}
\end{minipage}\hfill
\begin{minipage}[t]{0.48\linewidth}
\accessiblecodecaption{Rust}{lst:rust}
\begin{Verbatim}[commandchars=|\[\],fontsize=\footnotesize]
|codecm[/* code snippet */]
|codekw[let] name = |codestr["World"];
|codefunc[println!](|codestr["Hello, {name}!"]);
\end{Verbatim}
\end{minipage}\par\medskip

\noindent\begin{minipage}[t]{0.48\linewidth}
\accessiblecodecaption{C++}{lst:cpp}
\begin{Verbatim}[commandchars=|\[\],fontsize=\footnotesize]
|codecm[/* code snippet */]
|codekw[std::string] name = |codestr["World"];
|codefunc[std::cout] << |codestr["Hello, "] << name << |codestr["!"]
          << |codekw[std::endl];
\end{Verbatim}
\end{minipage}\hfill
\begin{minipage}[t]{0.48\linewidth}
\accessiblecodecaption{Scala}{lst:scala}
\begin{Verbatim}[commandchars=|\[\],fontsize=\footnotesize]
|codecm[/* code snippet */]
|codekw[val] name = |codestr["World"]
|codefunc[println](|codestr[s"Hello, $name!"])
\end{Verbatim}
\end{minipage}\par\smallskip

\begin{figure}
  \centering
  \begin{minipage}[t]{0.48\linewidth}
  \begin{Verbatim}[commandchars=|\[\],fontsize=\footnotesize]
name = |codestr["World"]
|codefunc[print](|codestr["Hello, "] + name + |codestr["!"])
\end{Verbatim}
\textbf{(a) Concatenation}
  \end{minipage}\hfill
  \begin{minipage}[t]{0.48\linewidth}
  \begin{Verbatim}[commandchars=|\[\],fontsize=\footnotesize]
name = |codestr["World"]
|codefunc[print](|codestr[f"Hello, {name}!"])
\end{Verbatim}
\textbf{(b) Interpolation}
  \end{minipage}
  \caption{The same output built by string concatenation (a) and by literal string interpolation (b) in Python. The central contrast studied in this paper is between these two forms.}
  \Description{Two Python code snippets side by side. The left snippet builds the greeting with the plus operator; the right snippet builds the same greeting with an f-string placeholder.}
  \label{fig:side-by-side}
\end{figure}


Although it can be tempting to argue based on code snippets (Figure~\ref{fig:side-by-side} and Listings~\ref{lst:java}--\ref{lst:scala}) that they are good or bad in some sense, \citet{10.1145/2534973} found that perceived intuition may be a function of experience in the language. In other words, a claim by an experienced programmer that something is intuitive does not make it so. Moreover, there is no known basis for these syntax variations across different languages. To the best of our knowledge, no formal study has been conducted to ascertain the effects of these syntax variations on program readability and comprehension. Previous studies have indicated that, on average, developers learn at least six different programming languages~\cite{10.1145/2509136.2509515}. However, there is clear evidence that developers find it difficult to learn more~\cite{10.1145/3377811.3380352,10.1145/2661136.2661156}. 

We found that string concatenation, particularly in complex expressions, significantly reduced program readability and increased comprehension time. Participants took longer on concatenation tasks than on interpolation tasks. A repeated-measures analysis of variance revealed a significant main effect of \textit{task category}, $F(1, 5026) = 92.35$, $p < .001$, $\eta^2_G = .018$, indicating differences in completion time between concatenation and interpolation tasks. There was also a smaller but statistically significant main effect of \textit{academic level}, $F(6, 5026) = 2.41$, $p = .025$, $\eta^2_G = .003$. The interaction between task category and academic level was not significant, $F(6, 5026) = 0.03$, $p = 1.00$, $\eta^2_G = .00$, suggesting that the effect of task category was consistent across levels of academic experience. Accuracy was generally high, with a moderate positive correlation between participant and correct responses (\(r = .55, p < .001\)), indicating consistent task understanding. Qualitative findings showed that participants viewed interpolation as cleaner and more intuitive, whereas concatenation was preferred mainly due to familiarity. Readability perceptions were shaped by experience and context, with participants emphasizing the importance of clear formatting, variable highlighting, and IDE support, indicating that readability depends as much on tooling and conventions as on syntax.

\subsection{Review of Relevant Scholarship}
There is a noticeable lack of formal scientific studies on string interpolation and concatenation within current research venues. We employed a combination of snowballing techniques and systematic literature search on major CS publication venues (conferences and journals) e.g. EMSE, ACM-TOCE, ICSE, ICER, ASE, SPLASH, POPL, ICFP, SIGCSE, ICPC, SCAM, ICSME, among others, from the year 01-01-2018 to 04-29-2024 using the following keywords: \textit{``string interpolation''}, \textit{``string concatenation''}, \textit{``string interpolation in computer programming''}. While strings are a common data type, we did not find any relevant evidence-based study on this topic. Therefore, we broadened our search parameters to include subjects associated with programming language constructs and their effects on program readability and comprehension. We also summarize other historical work in the area beyond our search.

Numerous studies have investigated the impact of various programming language constructs on program readability and comprehension, covering a wide spectrum of topics, addressing how programmers handle compile errors~\cite{osti_10446714,10.1145/3335814}, examined the readability of error messages across several studies~\cite{10.1145/3545945.3569791, 10.1145/2839509.2844584, 10.1145/2994154, 10.1145/3441636.3442320, 10.1145/3341525.3387384,10.1145/3411764.3445696}, explored comprehension of different programming constructs~\cite{hanenberg2022two, zid2024study}, and analyzed the programming approaches of novice programmers~\cite{10.1145/3626252.3630968, 10.1145/2676723.2677258, 10.1145/2994154}. This extensive body of research provides valuable insights into how language features influence coding practices and programmer interactions. \citet{10.1145/3545945.3569791} studied predicting the readability of programming error messages using previously developed readability factors as predictors. They trained machine learning models using a dataset of C and Java error messages. The results validate the previously proposed readability factors, and the models classify messages similarly to human raters. \citet{10.1145/3377811.3380352} conducted an analysis on Stack Overflow, targeting questions related to 18 programming languages. The research was driven by the hypothesis that existing knowledge might hinder the process of learning a new programming language. Through an examination of 450 specific queries on Stack Overflow, the research identified 276 (61.3\%) cases where developers' prior understandings from different programming languages led to incorrect assumptions and misunderstandings of a given help question on Stack Overflow.

Another interesting study by~\citet{hanenberg2022two} investigated readability differences between anonymous inner classes and lambda expressions in Java code snippets using two so-called N-of-1 trial studies. To be clear, the meaning of N-of-1 intended by the authors did not mean a single subject, as the name implies, but instead that the programming language tasks in question are so small that they can be repeated thousands of times by a small group to estimate the size of an effect. Put another way, it is a technique for piloting in the small. We mention it here because our study uses a similar approach, but scaled to a much larger sample and on a different topic.

In regard to the results, based on 9,600 observations, \citet{hanenberg2022two} found that lambda expressions without type annotations require less reading time, resulting in a reduction of reaction time of at most 35\%. Whether this result has an impact on daily programming activities is not yet clear in the literature; we view our work as a step toward answering analogous questions for string construction. Additionally, \citet{10.1145/3368089.3409701} and~\citet{uesbeck_et_al:OASIcs.PLATEAU.2018.1} investigated the effect of embedded programming language switching, also known as ``polyglot programming'', on productivity (development time). This study was motivated by previous research indicating that 97\% of open-source projects use more than one programming language and developers learn 6 languages on average, and 10 overall~\cite{10.1145/2544173.2509515, 10.1145/2745802.2745805}. Using a randomized controlled trial with repeated measures and a double-blind study design, the authors observed that the implications of polyglot programming on productivity are intricate and vary depending on the individual's level of experience. 

A recent study on functional program constructs' understandability was conducted by~\citet{zid2024study}. The authors performed a randomized controlled trial experiment with 209 developers to determine the understandability of python functional constructs (lambdas, list comprehension, map, reduce and filters). The authors found that code snippets containing lambdas are easier to modify compared to their procedural alternatives. 

\citet{Kluthe2025} conducted one of the first randomized controlled trials examining how nomenclature affects the usability of scientific computing languages. Their study addressed long-standing inconsistencies in the naming conventions of statistical functions across \texttt{Matlab}, \texttt{Python}, and \texttt{R}, investigating how intuitive or ``meaningful'' function names could improve learnability and user preference. Using a between-subjects design (\(n = 118\)) with a weighted ranked-choice survey, the authors found that semantically transparent and structured names were ranked by a large margin. Those that clearly describe the purpose of a function (e.g., \texttt{CompareMeans} rather than \texttt{t.test}) were strongly preferred over existing, often abbreviated or eponymous labels. The effect of programming language familiarity was negligible, while method naming exerted a statistically significant influence on user preference (\(\eta^2_p = 0.247\)). Related research has similarly emphasized the cognitive and educational consequences of language design and identifier naming. For example, \citet{Lawrie2006} and~\citet{Binkley2013} found that naming style (e.g., \texttt{camelCase} vs.\ underscores) and semantic consistency can affect program comprehension and accuracy. Whether this applies for strings in the way we investigate here is not clear.

Software-engineering research has long tried to pin down what makes code readable. \citet{Buse2010} built one of the first automated readability metrics, training a model on human-rated snippets and finding that low-level features like line length, identifier length, and the density of operators and parentheses track perceived readability well; the same study showed that this metric correlates with external signals of software quality such as defect reports. \citet{Scalabrino2016} added textual features to this model, such as whether comments and identifiers use consistent terminology and whether co-occurring terms are semantically related, and found that linguistic regularity matters as much as structural complexity. Later work has refined these models~\citep{Posnett2011, Scalabrino2021} and linked readability to downstream outcomes like maintenance and review behavior~\citep{Lenarduzzi2021}. Our study fits into this tradition. Concatenation and interpolation differ along several of the features these metrics flag as costly: operator density, punctuation, and how far a variable sits from the text it appears in. Comparing the two in a controlled setting lets us test whether the prediction implicit in this work, that visually denser and operator-heavier syntax slows comprehension down, holds for one of the most common syntactic choices in everyday programming.

We also explored publications not oriented towards research or academia. This includes computer programming blogs such as Medium~\footnote{\url{https://medium.com}}, Hacker News~\footnote{\url{https://news.ycombinator.com/}}, Stack Overflow~\footnote{\url{https://stackoverflow.com}}, and Stack Exchange~\footnote{\url{https://stackexchange.com}}, which are well-known among developers. These platforms are frequently used for discussions and technical writings on topics relevant to our study. We discovered several tutorials and articles on the topic of string interpolation. Nearly 90\% of these resources suggest that string interpolation offers an easy and concise syntax for embedding variables and expressions~\cite{stackoverflowInterpolation,mediumInterpolation,stackexchangeInterpolation}. However, none of these resources provide formal or informal evidence on how this method compares with concatenation. Furthermore, we observed an increasing demand from developers for the unification of syntax, especially in relatively new programming languages like Rust~\cite{discussion_rust}.

\subsection{Hypothesis, Aims, and Objectives}
To conduct our experiment in a meaningful and scientific manner, we established research questions and developed hypotheses to test our assumptions. To this end, the study aims to address the following research questions:
\begin{itemize}
    \item[\textbf{RQ1:}] 
    How does embedding variables, expressions, and functions within string literals affect program readability?
    \begin{itemize}
        \item[\textit{H\textsubscript{0}:}] 
        \textit{Embedding variables, expressions, and functions within string literals does not affect program readability.}
        \item[\textit{H\textsubscript{1}:}] 
        \textit{Embedding variables, expressions, and functions within string literals affects program readability.}
    \end{itemize}

    \item[\textbf{RQ2:}] 
    Relative to a concatenation baseline, how does literal string interpolation influence programmers' perceived readability, comprehension speed, debugging ease, and error propensity across tasks of varying complexity and levels of prior familiarity?
    \begin{itemize}
        \item[\textit{H\textsubscript{0}:}] 
        \textit{There is no difference between interpolation and concatenation across all primary outcomes (readability, comprehension time/cognitive load, debugging ease, and error rate).}
        \item[\textit{H\textsubscript{1}:}] 
        \textit{Interpolation outperforms concatenation on at least one primary outcome (higher readability, faster comprehension/lower cognitive load, easier debugging, or fewer errors).}
    \end{itemize}
\end{itemize}

\section{Method}
\subsection{Inclusion and Exclusion}
Participants in this study were drawn from a diverse range of educational and professional backgrounds to capture varying levels of programming experience and expertise. The inclusion criteria comprised undergraduate students (freshmen, sophomores, juniors, and seniors), graduate/post-graduate students, faculty members and other industry developers (hereafter referred to as professionals). Including these groups ensured representation across different stages of programming maturity and development, thereby enabling a comprehensive assessment of readability perceptions across experience levels. Whether the sample generalizes is an arguable point, but our point is we tried to make it diverse given the literature's consistent finding that the student level matters. While at least arguably interesting, we excluded participants under the age of 18.

The study specifically targeted participants within computer science or closely related disciplines, as the experimental tasks required a working understanding of programming constructs and string manipulation. Consequently, non-computer science majors and faculty were excluded to maintain the validity of the findings and prevent confounding effects arising from differences in domain expertise. 

\subsection{Participant Characteristics}
As noted earlier, participants were recruited from various educational stages at the University of Nevada Las Vegas (UNLV), including freshmen, sophomores, juniors, seniors, graduate students, and faculty members. All participants possessed prior experience with at least one programming language (e.g. C++), ensuring baseline familiarity with programming syntax and logic structures relevant to the study's tasks. While most participants were recruited directly through the UNLV department of Computer Science group email, additional recruitment efforts were made through a public broadcast shared on LinkedIn, which expanded outreach beyond the immediate university network. This external dissemination enhanced participant diversity and increased representation from individuals with varying academic and professional experiences in computer science. We provide the detailed participant characteristics in Table~\ref{tab:demographics}.

\begin{table}[htbp]
  \caption{Participant demographics: year in college, gender, and programming experience (n=314).}
  \label{tab:demographics}
  \centering
  \setlength{\tabcolsep}{6pt}
  \sisetup{table-number-alignment=right, table-text-alignment=right}
  \tagpdfsetup{table/header-rows={2}}
\begin{tabular}{l S[table-format=3] l S[table-format=3] l S[table-format=3]}
    \toprule
    \multicolumn{2}{c}{Participants} &
    \multicolumn{2}{c}{Gender} &
    \multicolumn{2}{c}{Programming Experience (years)} \\
    \cmidrule(lr){1-2}\cmidrule(lr){3-4}\cmidrule(lr){5-6}
    Group & {n} & Group & {n} & Years & {n} \\
    \midrule
    Freshman        & 46  & Female              & 66  & less than a year & 37 \\
    Sophomore       & 98  & Male                & 236 & 1                & 99 \\
    Junior          & 99  & Other               & 3   & 2                & 68 \\
    Senior          & 59  & Prefer not to say   & 9   & 3                & 55 \\
    Graduate        & 5   &                     &     & 4                & 35 \\
    Post-graduate   & 2   &                     &     & 5 and Above      & 20 \\
    Not Applicable  & 5   &                     &     &                  &    \\
    \midrule
    \textbf{Total}  & \textbf{314} &  & \textbf{314} &  & \textbf{314} \\
    \bottomrule
  \end{tabular}
\end{table}

Table~\ref{tab:demographics} summarizes the demographic composition of our study participants ($n=314$) across three dimensions: year in college, gender, and self-reported programming experience. The sample reflects a broad cross-section of students, with the largest groups represented by sophomores ($n=98$, 31.2\%) and juniors ($n=99$, 31.5\%). Freshmen ($n=46$, 14.6\%) and seniors ($n=59$, 18.8\%) were also well represented, while only a small number of participants identified as graduate ($n=5$) or post-graduate students ($n=2$). Five participants indicated that year-in-college information was not applicable to their situation.

In terms of gender, the sample had a larger proportion of people that identified as male ($n=236$, 75.2\%), with female participants comprising 21.0\% ($n=66$). A small number of participants identified as ``Other'' ($n=3$, 1.0\%) or preferred not to disclose their gender ($n=9$, 2.9\%). While the distribution skews toward male participants, the representation of female and non-binary participants ensures some diversity in perspectives.

Programming experience was distributed across a wide range of self-reported years. Thirty-seven participants (11.8\%) reported less than one year of experience, while the largest single group ($n=99$, 31.5\%) reported one year of experience. The remaining participants were distributed across two years ($n=68$, 21.7\%), three years ($n=55$, 17.5\%), four years ($n=35$, 11.1\%), and five or more years ($n=57$, 18.2\%). This spread indicates that while many participants were relatively early in their computing education, a substantial fraction had accumulated multi-year programming experience.

Overall, the demographic profile suggests that the participant pool was composed primarily of undergraduate students with varied but generally limited programming experience. This distribution is consistent with populations commonly studied in empirical software engineering research conducted in educational settings, and it provides an appropriate basis for analyzing the effects of educational interventions and programming practices in novice-to-intermediate contexts.

\subsection{Sampling Procedures}
This study employed purposive sampling~\cite{Tongco_2007}, a non-probability sampling technique commonly used in empirical software engineering research to intentionally select participants who possess relevant expertise or characteristics aligned with the study's objectives. In this case, participants were primarily computer science students, faculty members and industry professionals, all of whom had varying levels of programming experience. While the readability of source code can be of interest across multiple disciplines, the study focused on participants within computer science to establish a controlled baseline of understanding grounded in a shared technical background. This approach ensured that all participants could meaningfully interpret the experimental materials and engage with the coding constructs under investigation. Data collection and recruitment activities took place between \textbf{December 13, 2024}, and \textbf{June 30, 2025}. 

\subsection{Sample Size, Power, and Precision}
Power analysis is a standard procedure for determining an appropriate sample size in empirical studies. To the best of our knowledge, no prior research closely aligns with the methodological and experimental design conventions of the present work. Therefore, the expected effect size could not be estimated from the literature, and a conventional a priori power analysis based on comparable prior findings was not feasible.

To determine a target sample size, we conducted an a priori analysis for the primary within-subjects
comparison (Concatenation versus Interpolation), modeled as a two-tailed paired $t$-test with
$\alpha = .05$ and a target power of $.80$. Given the absence of prior benchmarks and the anticipated
variability inherent in human-subject program comprehension experiments, we adopted a conservative
small-effect assumption ($d = 0.20$), which yields a minimum required sample of $n = 199$. For
reference, a small-to-medium effect ($d = 0.30$) requires $n = 90$, and a medium effect ($d = 0.50$)
requires only $n = 34$.

We set a target of approximately 280 participants, above the small-effect minimum, to provide a buffer
against anticipated attrition and to ensure adequate precision for secondary subgroup analyses,
particularly comparisons across the seven academic-level groups. Recruitment exceeded this target,
yielding a final sample of $N = 314$. The complete power analysis code, including all assumptions and
post-hoc calculations, is available in the replication package~\cite{replication}. Beyond the formal power analysis, we also conducted a series of small pilot studies, described in Section 3.1, to refine task design and validate the data-logging mechanisms prior to main data collection.

\subsection{Measures and Covariates}
The primary objective of the experiment was to assess participants' task completion time and response correctness across different code comprehension scenarios. These measures were designed to evaluate whether variations in literal string interpolation syntax influence the readability and interpretability of program code. To establish a meaningful comparison, string concatenation was included as a baseline condition, allowing for a direct evaluation of readability differences between traditional concatenation methods and modern embedding approaches. Measuring both speed and accuracy enabled a balanced assessment of cognitive load and comprehension effectiveness under each experimental condition. Participants were explicitly informed that the purpose of these measurements was to examine the impact of syntactic style on readability, rather than to evaluate their individual programming proficiency. This clarification was intended to reduce potential performance anxiety and ensure authentic engagement with the tasks.

\subsection{Data Collection}
\begin{figure}
    \centering
    \includegraphics[width=\linewidth,alt={Flowchart of the study procedure from the pre-survey through tasks and post-survey.}]{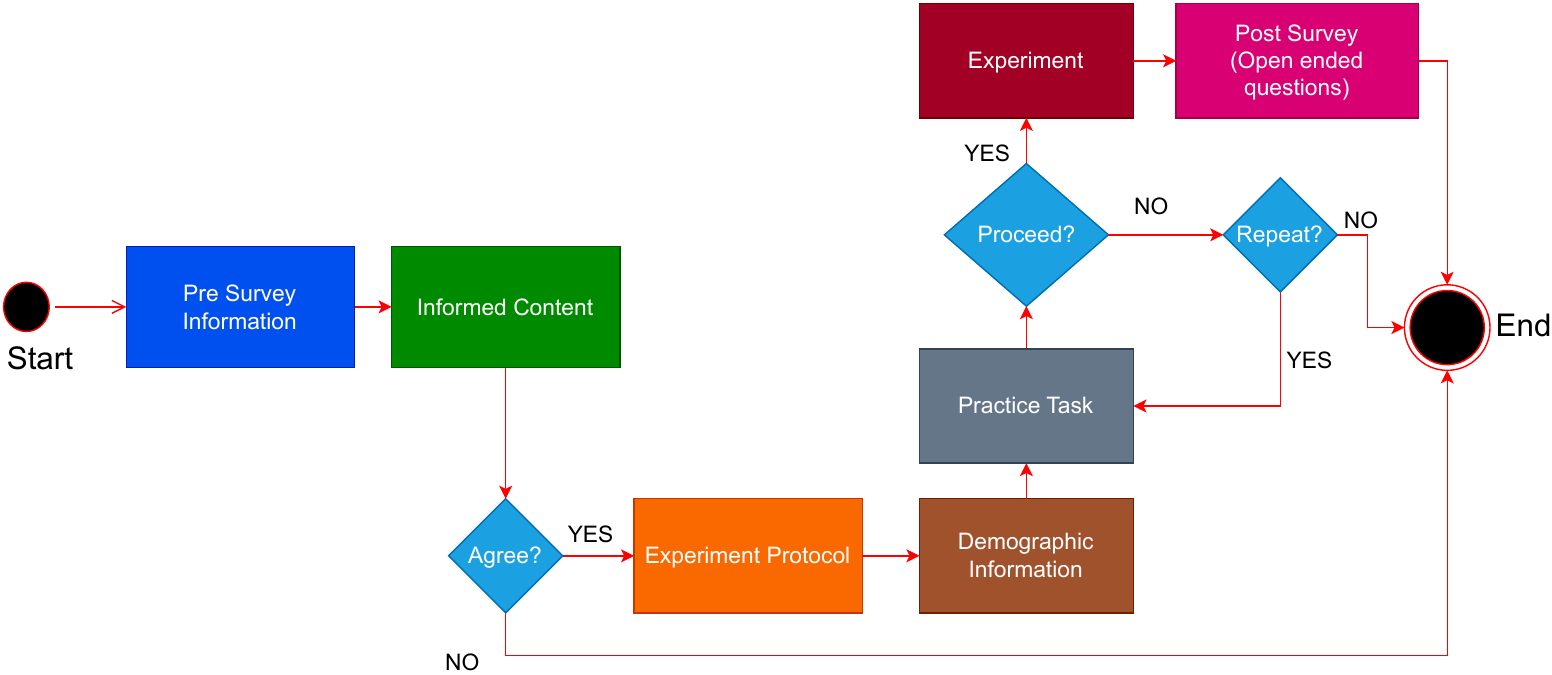}
    \caption{Overview of the experimental procedure showing the participant flow from pre-survey to post-survey stages. The process includes informed consent, practice tasks, main experimental trials, and open-ended post-experiment questions.}
    \Description{Overview of the experimental procedure showing the participant flow from pre-survey to post-survey stages.}
    \label{Fig:Tool}
\end{figure}

To collect experimental data, we developed a custom data collection tool designed specifically for this study. Figure~\ref{Fig:Tool} illustrates the overall experimental flow, outlining the sequence from participant recruitment and consent to practice tasks, experimental trials, and the post-survey phase that captured qualitative feedback on code readability and comprehension. The tool first prompted participants to complete a brief demographic form (cf. Figure~\ref{Fig:Demographics}), which gathered essential background information such as educational level, programming experience, and familiarity with string manipulation. These data were later used during the analysis phase to explore potential relationships between participant characteristics and performance outcomes. Before beginning the main experiment, participants completed four practice tasks. These practice rounds were included to help participants become familiar with the experimental interface, timing, and overall procedure. This step was important to ensure that participants felt comfortable with the task format and could focus on comprehension rather than navigation or mechanics.

After submitting their demographic information, participants could start the practice phase by pressing any key or clicking the ``Next'' button. The tool allowed participants to repeat the practice tasks as many times as needed. Once they felt confident, they could proceed to the main experiment by pressing the ``Proceed'' button or hitting the Enter/Return key. All experimental results were automatically saved in a comma-separated values (CSV) file following the tidy data format, ensuring that each variable was stored in a column and each observation in a row. This structure facilitated clean and consistent data analysis in subsequent stages of the study.

\subsection{Quality of Measurements}
All measurements in this study were automatically recorded by the experimental platform, which helped minimize human error and ensure consistent data collection across participants. The system captured two key variables: \textit{completion time (in seconds)} and \textit{response correctness}. To promote measurement accuracy and participant confidence, the experiment included four practice tasks that could be repeated as many times as desired before beginning the main session. This familiarization process allowed participants to understand the task structure, timing, and interaction flow, thereby reducing potential confusion or learning effects during the actual experiment. In addition, the platform incorporated several input validation mechanisms and keyboard safeguards to prevent invalid responses or accidental submissions. These features ensured that every recorded data point reflected a deliberate and valid participant action, thereby strengthening the reliability and overall quality of the collected measurements. Another way to think about these training periods is we were trying to avoid novelty effects due to the participants being unfamiliar with our experimental system itself.

\subsection{Instrumentation}
Figure~\ref{Fig:DataCollectionTool} illustrates the custom-built experimental platform used to administer the study. As described earlier, this platform was developed by the authors to enable efficient data collection and maintain precise control over all experimental conditions. The system automatically presented tasks, captured participants' responses, and recorded timing data with high accuracy, thereby ensuring consistency and reliability across all sessions. To demonstrate the task structure, we present two examples, one for the concatenation condition and another for the interpolation condition. These examples highlight how participants interacted with the platform and the nature of the comprehension tasks they were required to perform. The full set of experimental tasks is included in the Appendix and made publicly available through the replication package~\cite{replication}. Table~\ref{Tab:TaskComplexity} provides a summary of the task complexity levels used in the experiment. Each level was defined by the number of variables embedded within a string expression, allowing for a controlled increase in cognitive demand across tasks.
\begin{table}
  \caption{Level of Task Complexity}
  \label{Tab:TaskComplexity}
  \renewcommand{\arraystretch}{1.3}
  \tagpdfsetup{table/header-rows={1}}
\begin{tabular}{llp{8cm}} 
    \toprule
    \textbf{Level} & \textbf{Complexity} & \textbf{Description} \\
    \midrule
    Level 1 (L--1) & Low & Tasks that involve concatenation or interpolation of \textbf{two} variables. \\
    Level 2 (L--2) & Moderate & Tasks that involve concatenation or interpolation of \textbf{three} variables. \\
    Level 3 (L--3) & High & Tasks that involve concatenation or interpolation of \textbf{four} variables. \\
    Level 4 (L--4) & Very High & Tasks that involve concatenation or interpolation of \textbf{more than four} variables. \\
    \bottomrule
  \end{tabular}
\end{table}

\subsection{Masking}
To maintain a reasonable level of masking, we did not disclose the hypothesis of the study to the participants and we carefully controlled the format of the tasks to be as neutral as we reasonably could. However, masking has inherent limitations in studies like ours. Unlike other fields, computer programming experiments require participants to closely observe the code snippets being studied \cite{10.1145/3291279.3339421}. As a result, it is conceivable that participants might deduce the underlying hypothesis. Moreover, we recruited from different groups who might be well versed with interpolation and concatenation constructs. Nonetheless, feedback received from several participants indicated that they were not aware of the original hypothesis; instead, their attention was primarily focused on providing accurate responses to each task.

\subsection{Psychometrics}
We could not find any known reliability coefficient that can be used as a benchmark for metrics we are measuring. However, we evaluated reliability using the test–retest method, which assesses how consistently a measure performs over time~\cite{Guttman1945}. To do this, we conducted an N-of-1 trial \cite{GUYATT199088}, where a single participant completed the same set of tasks on two different days. Each session produced 100 observations, allowing us to compare performance stability across time. We then calculated the Pearson correlation coefficient (\texttt{r}) to determine the degree of consistency between the two sessions (S1 and S2). The analysis revealed a strong positive correlation, \texttt{r(2) = 0.73, p = 0.017}, indicating that participants' performance remained stable across repeated measurements. Our point is that this is hardly sufficient for a full psychometric breakdown of our measures, but we at least checked if doing the same tasks gives us the same answers as a starting point.

\subsection{Conditions and Design}

This study employed a within-subjects repeated-measures design with two primary dependent variables:
(i) task completion time and (ii) response correctness. Each participant completed tasks under both experimental conditions (i.e. string concatenation and string interpolation), allowing direct comparison of performance across conditions while controlling for individual differences. The experiment followed an N--of--1 trial design, inspired by the methodological framework of~\citet{hanenberg2022two}. This approach allowed for fine-grained analysis of participant performance over repeated exposures to systematically varied programming tasks. Using the custom-built experimental platform, random code snippets were dynamically generated for each trial. Each task presented a short code segment and four multiple-choice options, from which participants selected the correct output or interpretation.

\textbf{Task Generation.} 
Task instances were generated automatically using a template-based generator. Each of the 16 tasks was instantiated from a fixed template defined by its structural pattern, complexity level (L--1 to L--4), and string construction mechanism (concatenation or interpolation). For each task, the concatenation and interpolation variants were generated from the same template instantiation, differing only in the string construction mechanism to ensure a controlled comparison. Distractor options were created through predefined perturbation rules while enforcing a single correct answer. The templates were piloted before the main study, and ambiguous items were revised (Section~3.1). The replication package includes the complete set of templates, perturbation rules, and all task instances used in the study.

To separate the conceptual contrast between concatenation and interpolation from familiarity with any one language's syntax, we presented all stimuli in a language-neutral abstract notation: ``\texttt{+}'' denotes concatenation and ``\texttt{\{~\}}'' denotes interpolation, with each operator's meaning stated in the task instructions. Participants were not asked to execute the snippets but to reason about the resulting string at the conceptual level. This follows established practice in computing-education and program-comprehension research, where pseudo-language stimuli reduce variation from participants' differing language backgrounds. \citet{Tew2011}, for instance, use a pseudo-language to assess CS1 knowledge across language backgrounds, and \citet{10.1145/2534973} adopt similar abstractions in their work on syntax perception. Holding the surface syntax constant reduces the risk that background fluency, for example in Python's f-strings or Java's \texttt{String.format},
confounds the comparison.

\begin{llmprompt}[Task 7 illustrating a concatenation-based code comprehension question.]{Fig:TaskConcatenation}
\small
\textbf{Instruction:} The ``+'' symbol represents concatenation. Press 1, 2, 3, or 4 to select the correct concatenation.

\medskip
\textbf{Code Snippet:}
\begin{verbatim}
"World, " + 1 + 2 + 3 + (1 + 1 + 1 + 1) + ("5" + (6 + 7))
\end{verbatim}

\textbf{Options:}
\begin{enumerate}
  \item ``\verb|World, 123("5"(67))|''
  \item ``\verb|World, 123("5"))|''
  \item ``\verb|123(1111)("5"(67)), World|''
  \item ``\verb|World, 123(1111)("5"(67))|''
\end{enumerate}

\textit{Correct answer: option (4). Evaluating left to right yields World, 123(1111)("5"(67)); parentheses are emitted as literal characters.}
\end{llmprompt}

\begin{llmprompt}[Task 8 illustrating an interpolation-based code comprehension question.]{Fig:TaskInterpolation}
\small
\textbf{Instruction:} The ``\{ \}'' symbol represents interpolation. Press 1, 2, 3, or 4 to select the correct interpolation.

\medskip
\textbf{Code Snippet:}
\begin{verbatim}
"{1} {2} {3} {({9} {9} {9} {1})} World { ({"5"} {(67)} )}"
\end{verbatim}

\textbf{Options:}
\begin{enumerate}
  \item ``\verb|1 2 3 (9 9 9 1) World ("5" (67) )|''
  \item ``\verb|123(9991)World("5"(67))|''
  \item ``\verb|1 2 3 ({9} {9} {9} 1) World ("5" (67) )|''
  \item ``\verb|1 2 3 9 9 9 1 World "5"67|''
\end{enumerate}

\textit{Correct answer: option (1).}
\end{llmprompt}

Tasks were randomized and alternated between the concatenation and interpolation conditions to minimize order effects. For clarity and consistency, the symbols ``+'' and ``\{ \}'' were used to represent concatenation and interpolation, respectively. Boxes~\ref{Fig:TaskConcatenation} and~\ref{Fig:TaskInterpolation} illustrate example tasks used in the experiment, demonstrating the concatenation and interpolation conditions, respectively. The complete list of task items and their corresponding code snippets are provided in the Appendix~\ref{appendix:taskslist}. During each trial, participants were required to read the presented code snippet and select their response by pressing the numeric keys ``1'', ``2'', ``3'', or ``4'' on the keyboard. Once a response was made, it could not be changed, and the next task appeared automatically. The experiment as a whole was not time-limited; however, completion time and correctness were automatically recorded for each individual task. All results were stored in a CSV file following the tidy data format. Participants were not allowed to revisit or modify previous answers, ensuring that each response reflected their immediate comprehension and interpretation of the presented code.

\subsection{Data Diagnostics}
The experimental platform used for data collection was designed in accordance with the principles of tidy data~\cite{JSSv059i10}, ensuring that the resulting dataset was well-structured, consistent, and free of missing values. Because the study focused on two direct measurements, task completion time and response correctness, no additional data transformations were necessary. During preliminary quality checks, we identified a single case of data duplication that originated from the experiment platform. This issue was resolved by removing the duplicate entry and retaining only one valid record. After this correction, the final dataset comprised 314 participants. Because the study aimed to capture the full range of participant behavior, outliers were not excluded. Instead, they were treated as meaningful observations that might reflect authentic variation in programming comprehension and task performance. To identify potential outliers, we applied a conventional interquartile range (IQR) criterion~\cite{tukey1977exploratory}. Specifically, observations falling below $Q1 - 1.5 \times IQR$ or above $Q3 + 1.5 \times IQR$ were flagged as potential outliers. Although flagged, these observations were retained in all subsequent analyses to preserve the natural variability inherent to human-performance data and to avoid artificially narrowing the distribution of completion times. These outliers were carefully examined and are discussed in Section~\ref{sec:discussion}.

\subsection{Analytic Strategy}

To evaluate the effects of string interpolation and concatenation on code readability, we employed repeated-measures inferential statistical analyses that account for within-subject variability. Because the same participants completed tasks under multiple conditions, all comparisons were analyzed using repeated-measures analysis of variance (ANOVA), allowing us to assess differences across conditions while appropriately modeling correlated observations. For factors involving more than two levels, such as task complexity, repeated-measures ANOVA was used to examine systematic changes in performance across conditions.

To ensure the validity of the repeated-measures analyses, we assessed the assumption of sphericity using Mauchly's test~\citep{Mauchly1940}. When the assumption of sphericity was violated, Greenhouse–Geisser corrections were applied to the degrees of freedom, as recommended for repeated-measures ANOVA~\citep{girden1992anova}. Effect sizes are reported using generalized eta squared ($\eta^2_G$), which provides a conservative and comparable estimate for within-subject designs. This adjustment ensured that multiple comparisons were interpreted conservatively, thereby enhancing the reliability and validity of the statistical conclusions. In addition to significance testing, we computed effect size measures to quantify the magnitude of observed differences. For repeated-measures analyses, general eta-squared ($\eta_G^2$) was reported to indicate the proportion of variance explained by each factor.

\subsection{Qualitative Analysis}\label{sec:qual}
To address RQ2, we conducted a qualitative analysis of participants' responses to five open-ended post-survey questions focusing on: (1) readability rationale, (2) comprehension and debugging, (3) learning curve, (4) preference, and (5) improvement suggestions. The purpose of this analysis was to gain deeper insights into how participants perceived and reasoned about the readability and usability of code written using literal string interpolation and string concatenation. In particular, we sought to understand how these two syntactic styles influenced participants' perceptions of readability, comprehension speed, debugging ease, and error propensity across tasks of varying complexity and levels of prior familiarity.

We adopted the Framework Method~\cite{Gale2013}, which involves familiarization with the data, open coding, developing and applying an analytical framework, charting data into a case-by-code matrix, and interpreting themes. To reduce the substantial manual effort typically required for open coding, we incorporated large language models in line with recent guidelines for qualitative analysis in empirical software engineering~\cite{baltes2025guidelines,BanoMuneera,11071418}. In our study, the LLM supported the \emph{coding stage} by generating draft codes, which were then refined and structured by human researchers. This integration preserved efficiency while ensuring rigor and interpretability. \citet{sserunjogi2026prompt} similarly found that LLM-assisted annotation reproduced human judgments unevenly across coding dimensions, motivating a hybrid human--LLM annotation strategy like the one used here.

We leveraged ChatGPT (version 5.0)~\cite{chatgptv5.0} to support and accelerate the initial open coding phase and promote consistent application of coding definitions. The model was run at default temperature, with the full response CSV for each question supplied per session. Prompts were developed iteratively on Q1 responses as a calibration set, which surfaced three failure modes and their fixes: overly broad codes (addressed by a five-step METHOD section requiring sub-themes), untraceable or paraphrased quotes (addressed by requiring verbatim quotes tagged with respondent IDs and verified against the sources), and inconsistent label length (addressed by a $\leq$4-word constraint). The manuscript reports the final prompt. Robustness was assessed by re-running it on three stratified 20\% subsets of Q1 responses (minimum theme-level Jaccard 0.86 across all pairwise comparisons) and on two paraphrased role/persona blocks, with no change in the resulting major themes.

The outputs generated by ChatGPT were carefully reviewed, validated, and refined by the authors, as described in the subsequent paragraphs. For each question, we supplied a standardized, well-structured prompt (see Box~\ref{prompt:llmprompt} for open-ended question 1) that required: (i) non-overlapping code labels ($\leq$4 words) with definitions and explicit \emph{when-to-use / when-not-to-use} criteria; (ii) at least one verbatim quote ($\leq$30 words) per code, labeled with respondent ID (R-\#); and (iii) explicit comparison to the concatenation baseline where relevant. Prompt versions, calibration samples, error logs, and the Jaccard computation script are in the replication package~\cite{replication}.

\begin{llmprompt}[Example ChatGPT Prompt for Qualitative Analysis of Open-ended Question 1]{prompt:llmprompt}
\small
\ttfamily

\textbf{ROLE / PERSONA}\\
You are a senior qualitative researcher specializing in empirical software engineering. You apply the Framework Method and grounded theory techniques to perform inductive open coding, card sorting, and theme synthesis from textual survey data.

\textbf{OBJECTIVE}\\
Analyze participant responses to the following question:
\begin{quote}\itshape
\textbf{``Based on your experience with both string interpolation and string concatenation in the survey, which approach did you find more readable, and why? Please provide specific examples or reasons to support your viewpoint.''}
\end{quote}
For each response, produce concise, human-interpretable qualitative codes that capture the respondent's reasoning about readability and comprehension.

\textbf{INPUTS (provided)}%
\begin{itemize}
  \item A CSV file containing a free-text column (\texttt{ResponseText}) with participant answers.
  \item Context: Responses come from developers with prior programming experience comparing readability between string interpolation and concatenation.
\end{itemize}

\textbf{METHOD (must follow)}%
\begin{enumerate}
  \item \textbf{Open Coding (Inductive):} Examine each response and identify key ideas related to readability, comprehension, debugging effort, visual clarity, familiarity, or contextual preference.
  \item \textbf{Code Labeling:} Assign non-overlapping \textbf{Theme} and \textbf{Sub-theme} labels (\(\leq 4\) words each) that best capture the meaning. Provide clear, distinct labels—avoid generic or overlapping terms.
  \item \textbf{Decision Rules:} For each code, briefly note when the code should or should not be applied (e.g., \emph{use when the participant attributes readability to syntax familiarity; not when the response discusses task complexity alone}).
  \item \textbf{Illustrative Quote:} Include one verbatim quote (\(\leq 30\) words) from the response, tagged with the respondent ID (e.g., R-142), that exemplifies the sub-theme.
  \item \textbf{Comparative Reference:} When applicable, indicate whether the statement compares interpolation to concatenation directly, implicitly favors one, or expresses contextual neutrality.
\end{enumerate}

\textbf{OUTPUT FORMAT (CSV; one row per response)}%
\begin{itemize}
  \item \textbf{Theme}: A concise label summarizing the main perspective (e.g., \emph{Interpolation Preferred}, \emph{Concatenation Familiarity}, \emph{Context Dependent}, \emph{No Clear Preference}).
  \item \textbf{Sub-theme(s)}: One or more detailed categories separated by commas (e.g., \emph{Readability Clarity}, \emph{Habit Influence}, \emph{Code Length Sensitivity}).
  \item \textbf{Quote}: A short verbatim example (\(\leq 30\) words) with the respondent ID (R-\#).
  \item \textbf{Notes}: Brief rationale or when-to-use guidance for human validation.
\end{itemize}

\textbf{ADDITIONAL OUTPUT (Structured Codebook Summary)}%
\begin{itemize}
  \item A concise table summarizing:
    \begin{enumerate}
      \item \textbf{Theme Name}
      \item \textbf{Definition} (what the theme captures)
      \item \textbf{When-to-Use} and \textbf{When-Not-to-Use} criteria
      \item \textbf{Example Quote} (R-\#)
    \end{enumerate}
  \item The codebook should be logically ordered (e.g., by theme frequency or conceptual hierarchy) and written in plain, human-readable language suitable for replication and refinement.
\end{itemize}
\end{llmprompt}

The outputs produced from these prompts were treated strictly as drafts. After the draft codebook and preliminary codes were produced, two researchers with prior experience in empirical software engineering independently double-coded a stratified subset (30\%) of responses. While we did not compute Cohen's Kappa due to the qualitative nature of the task, agreement was verified through iterative discussion and convergence on shared meanings. Analytic decisions and reflections were documented in memos to ensure reflexivity and transparency throughout the process. Insights from this validation informed refinement of the codebook, including merging overlapping categories, splitting ambiguous labels, and clarifying usage rules.

The refined codebook was then systematically applied across the entire dataset. Specifically, LLM draft annotations for all responses were re-examined under the updated definitions, with human analysts correcting misapplied codes and ensuring consistent application across all responses. We also tracked recurring LLM errors (e.g., overgeneralized categories, hallucinated labels, or omitted context) to ensure these issues were corrected consistently during refinement and transparently reported. This error summary is included in the replication package~\cite{replication}. In cases of hallucination, overgeneralization, or omitted context, the human-authored interpretation took precedence. This process ensured that every response was covered under a validated thematic structure rather than relying on unchecked LLM output. Coded excerpts were charted into a case-by-code matrix to facilitate systematic comparison across themes and participants, consistent with Stage 6 of the Framework Method. Using the refined codebook, we organized the data into themes and sub-themes, each supported by at least one verbatim quote (R-\#). All prompts, model settings, the final codebook, coded tables, agreement notes, and anonymized excerpts are included in the replication package to support clarity, reproducibility, and professional reporting.

\section{Results}
\vspace{0.5\baselineskip}
\begin{framed}
\underline{\textbf{RQ1:}} How does embedding variables, expressions and functions with string literals affect program readability?
    \begin{itemize}
        \item[\textit{H\textsubscript{0}:}] \textit{Embedding variables, expressions, and functions within string literals does not affect program readability.}
        \item[\textit{H\textsubscript{1}:}] \textit{Embedding variables, expressions, and functions within string literals affects program readability.}
     \end{itemize}
\end{framed}

\subsection{Participant Flow}\label{sec:participant-flow}
Prior to formal data collection, the experimental protocol underwent a pilot phase conducted by the first author and reviewed by the co-authors to refine the study design and procedures. The pilot employed an N-of-1 trial design comprising approximately 100 repeated trials, which enabled calibration of task difficulty, verification of timing accuracy, and assessment of test–retest reliability.

Insights from the pilot phase informed several refinements to the final study design, including adjustments to task complexity levels, clarification of task instructions, and validation of the data-logging mechanisms used to capture completion time and accuracy. No participant data from the pilot phase were included in the final analyses. The finalized protocol was then used consistently for all 314 registered participants in the main study.

\subsection{Recruitment}
Participant recruitment was conducted using a web-based, custom-built experimental platform designed specifically for this study. Invitations containing the study link were distributed through Discord, LinkedIn and Email, enabling participants to access the platform remotely and complete the experiment at their convenience. Recruitment occurred between December 13, 2024, and June 30, 2025. The study was reviewed and approved by UNLV Institutional Review Board (\textbf{No. UNLV-2024-339}). All participants were informed about the study's purpose, procedures, and their rights prior to participation, including the option to withdraw at any time without penalty. Participants affiliated with UNLV were compensated for their participation with course credit of up to 2\% in CS 472/672 - Software Product Design and Development I, CS 135 - Computer Science I, and CS 218 - Introduction to Systems Programming, CS 789 - Software Re-engineering.

\subsection{Statistics and Data Analysis}\label{sec:statistics}
\begin{figure}
  \centering
  \includegraphics[width=0.82\linewidth,alt={Boxplot comparing task completion durations in seconds across experimental task groups.}]{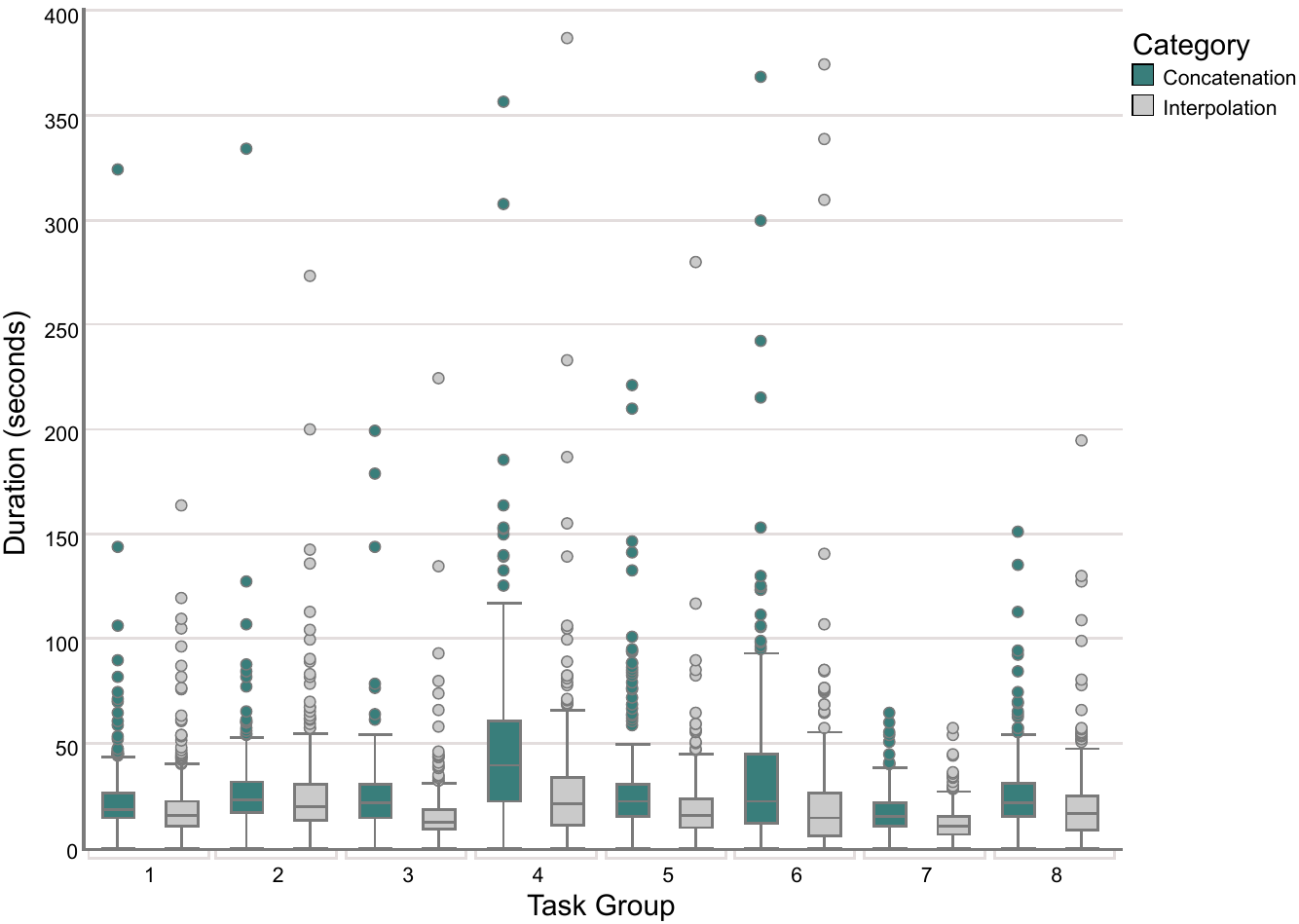}
  \caption{Distribution of task completion duration across task groups for both concatenation and interpolation categories. Each boxplot represents the spread of completion times (in seconds), with outliers shown as individual points.}

  \Description{Boxplot showing tasks completion time (in seconds)}
  \label{Fig:TaskGroup-Duration}
\end{figure}

In this section, we present results of our experiment, which was performed repeatedly to generate 314 data points from 16 tasks. We used Quorum programming language~\cite{10.1145/3017680.3022377} for the statistics and data analysis. Table~\ref{Tab:ExtendedTaskStats} shows the average (avg.) duration (in seconds) to complete each task and the standard deviation (std.). Overall, each task takes an average of $\approx$ 24.26 seconds with a standard deviation of $\approx$ 23.91.
\begin{table}[htbp]
\centering
\caption{Descriptive statistics of task completion time and accuracy across experimental tasks. Each row represents a unique task identifier (Task ID) with corresponding task category, complexity level, mean completion time ($M$), standard deviation ($SD$), median time, and accuracy rate.}
\label{Tab:ExtendedTaskStats}
\renewcommand{\arraystretch}{1.2}
\setlength{\tabcolsep}{4pt}
\tagpdfsetup{table/header-rows={1}}
\begin{tabular}{lccccc}
\toprule
\textbf{Task ID} & \textbf{Task Category} & \textbf{Complexity} & \textbf{$M \pm SD$ (s)} & \textbf{Median (s)} & \textbf{Accuracy (\%)} \\
\midrule
1  & Concatenation  & L--1 & 23.76 $\pm$ 22.92 & 20.10 & 91.5 \\
2  & Interpolation  & L--2 & 20.26 $\pm$ 18.47 & 17.35 & 93.0 \\
3  & Concatenation  & L--3 & 27.14 $\pm$ 23.49 & 25.02 & 88.7 \\
4  & Interpolation  & L--2 & 26.20 $\pm$ 26.71 & 23.80 & 90.2 \\
5  & Concatenation  & L--3 & 24.04 $\pm$ 19.69 & 22.10 & 89.4 \\
6  & Interpolation  & L--1 & 16.24 $\pm$ 17.89 & 14.00 & 94.8 \\
7  & Concatenation  & L--4 & 45.57 $\pm$ 39.23 & 40.50 & 81.3 \\
8  & Interpolation  & L--4 & 27.87 $\pm$ 33.66 & 23.60 & 86.9 \\
9  & Concatenation  & L--3 & 26.91 $\pm$ 26.64 & 25.11 & 88.0 \\
10 & Interpolation  & L--1 & 19.46 $\pm$ 21.35 & 17.08 & 93.6 \\
11 & Concatenation  & L--4 & 33.90 $\pm$ 40.36 & 28.20 & 82.5 \\
12 & Interpolation  & L--2 & 22.66 $\pm$ 36.96 & 20.33 & 91.1 \\
13 & Concatenation  & L--1 & 16.54 $\pm$ 10.83 & 15.45 & 94.5 \\
14 & Interpolation  & L--1 & 12.14 $\pm$ 8.83  & 11.02 & 95.8 \\
15 & Concatenation  & L--2 & 24.65 $\pm$ 19.38 & 22.99 & 90.0 \\
16 & Interpolation  & L--2 & 20.06 $\pm$ 20.37 & 18.21 & 92.7 \\
\midrule
\textbf{Overall Mean} & -- & -- & \textbf{24.26 $\pm$ 23.91} & \textbf{21.10} & \textbf{90.6} \\
\bottomrule
\end{tabular}
\end{table}
Figure \ref{Fig:TaskGroup-Duration} displays the completion times for each task in the form of a boxplot, showing variations in the completion times across tasks. Notably, Tasks 7, 8, 11 and 12 exhibit a broad variance compared to the other tasks. This observation is intriguing and will be discussed further in the subsequent section. We believe it suggests a difference in task categories (concatenation vs. interpolation) for each respective task.

\begin{figure}
  \centering
  \includegraphics[width=0.85\linewidth,alt={Comparison of completion times between string concatenation and interpolation tasks.}]{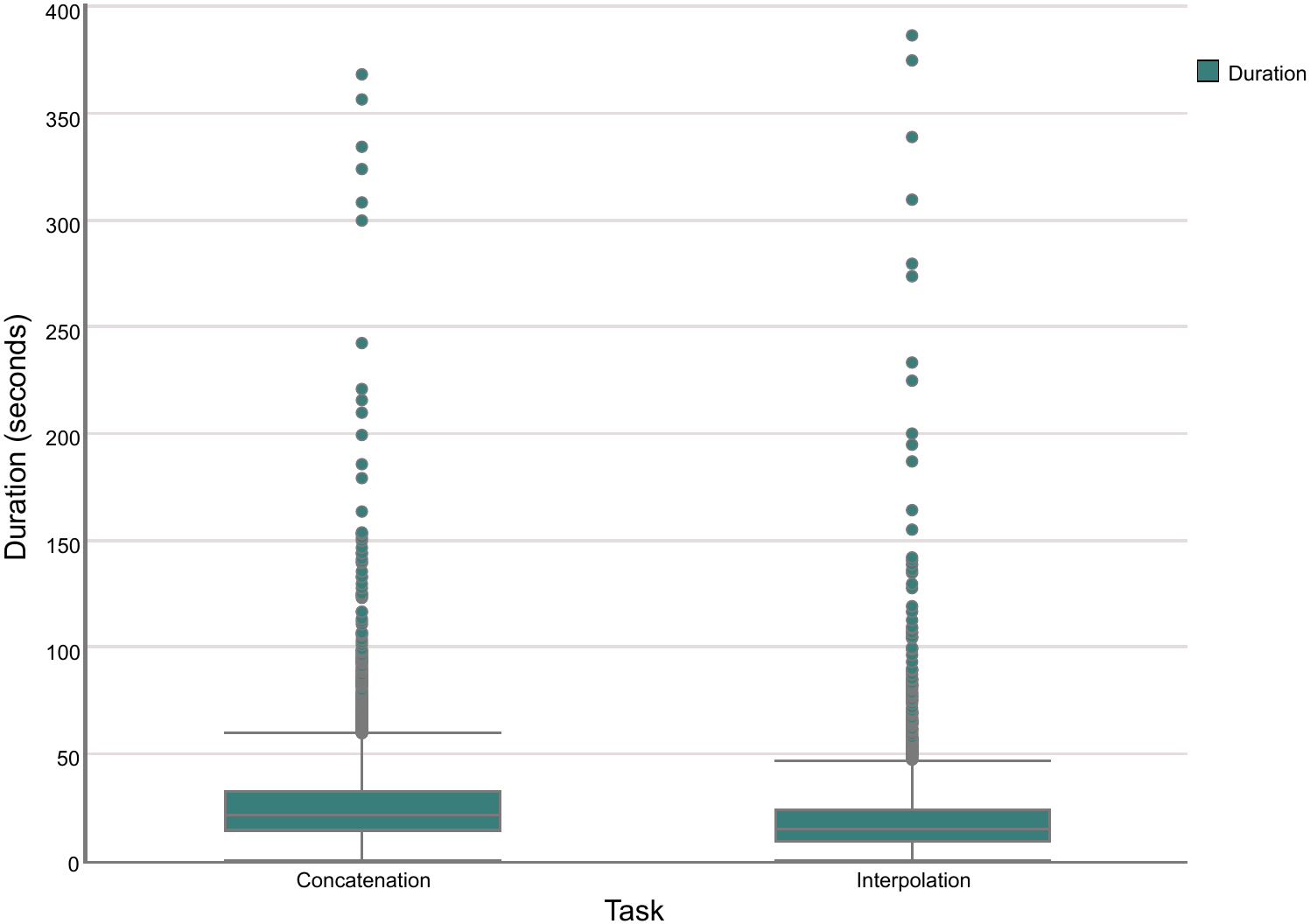}
  \caption{Comparison of overall task completion duration between concatenation and interpolation conditions. Each boxplot represents the distribution of completion times (in seconds), with interpolation tasks showing lower median durations and fewer extreme outliers compared to concatenation tasks.}
  \Description{Comparison of Task Completion Time Between the Groups}
  \label{Fig:BoxPlotOverall}
\end{figure}

\begin{figure}
  \centering
  \includegraphics[width=0.85\linewidth,alt={Comparison of task completion durations by academic year and string-construction condition.}]{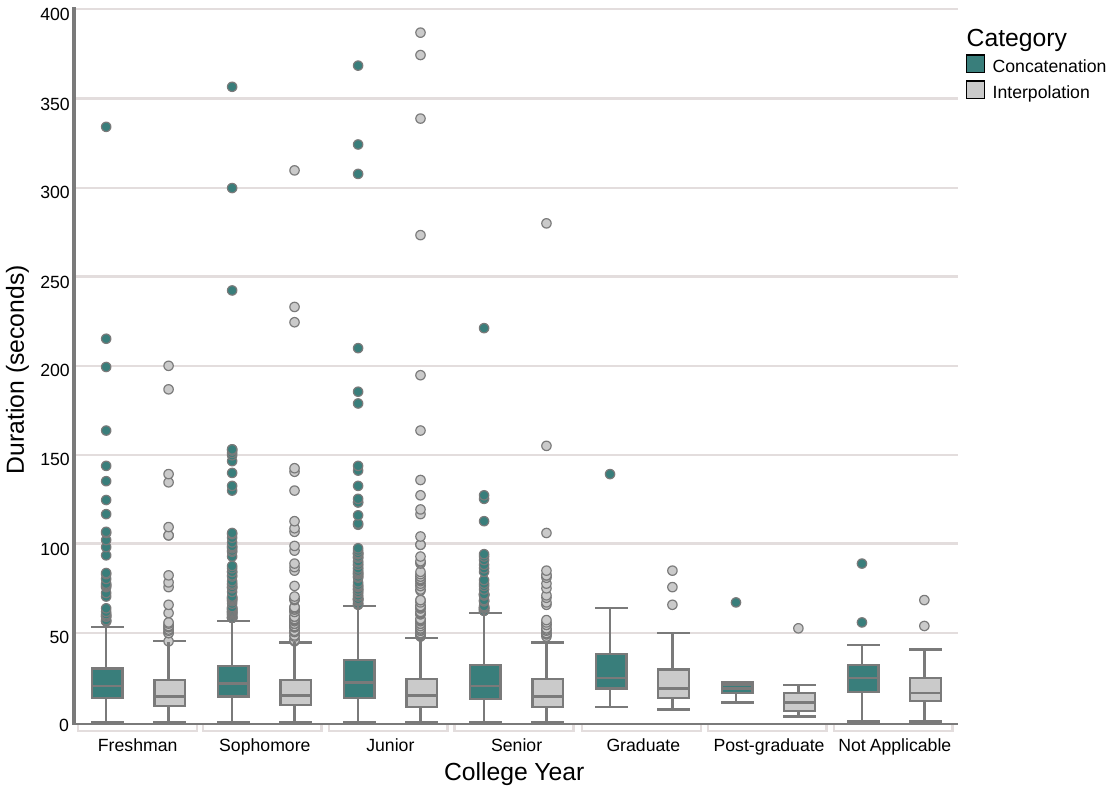}
  \caption{Distribution of task completion duration across college year groups for both concatenation and interpolation categories. Each boxplot represents the spread of completion times (in seconds) for participants at different academic levels, with outliers shown as individual points.}
  \Description{Comparison of College Year and Tasks Completion Time for Each Category}
  \label{Fig:YearInCollege-Duration}
\end{figure}

Figure \ref{Fig:BoxPlotOverall} presents the results of a comparison of task completion times between task categories. The two distributions overlap considerably, although interpolation tasks show a lower median and fewer extreme values. Figure~\ref{Fig:YearInCollege-Duration} further suggests some variation in completion time across college years. 

To test whether task completion time differed by \textit{task category} (concatenation vs.\ interpolation) and \textit{college year}, we conducted a repeated measures ANOVA with task category and year in college as a within-subject factors. The analysis revealed a significant main effect of task category, $F(1, 5026) = 92.35$, $p < .001$, $\eta^2_G = .018$, and a smaller but statistically significant main effect of year in college, $F(6, 5026) = 2.41$, $p = .025$, $\eta^2_G = .003$. The interaction between task category and year in college was not significant, $F(6, 5026) = 0.03$, $p = 1.00$, $\eta^2_G = .00$. Given the significant main effect of academic level, a post-hoc Tukey–Kramer test was conducted across all 21 pairwise comparisons of college year groups. Only one comparison reached significance: Juniors completed tasks faster than Seniors, $t(5033) = 2.99, p = .044$. All remaining 20 pairwise comparisons were non-significant ($ps = .253$--$1.0$; see replication package~\cite{replication} for full results). In light of the very small effect size ($\eta^2_G = .003$), this single significant contrast should be interpreted cautiously — it represents a marginal difference that accounts for less than 1\% of total variance in completion time and does not alter the primary conclusion that the effect of academic level on task performance was negligible.

We also acknowledge that academic level is an imperfect proxy for programming experience. Prior work, however, shows that self-reported years of programming experience is itself a noisy measure: \citet{Siegmund2014modeling} argue that programming experience is multi-dimensional and that simple year-counts often correlate weakly with task performance, and \citet{6240511} note that there is no agreed-upon way to measure programming experience, with researchers using a wide variety of proxies. More recent work reinforces this: \citet{10.1145/3236024.3236061} found that experience and expertise are only loosely related and that self-assessments are context-dependent, and \citet{6882243} argue that programming skill requires a validated instrument rather than a self-reported duration. 

We chose academic year because it captures structured curricular exposure and is a standard variable in computing education research. To verify that our findings do not depend on this choice, we re-fit our primary model as an ANCOVA (implemented as a linear mixed-effects regression) with self-reported years of programming experience as a continuous covariate. Programming experience showed a small but significant effect on completion time $(\beta = -0.42~\text{s/year}, p = .018)$, with more experienced participants completing tasks slightly faster. Controlling for programming experience did not alter the primary finding: the main effect of task category remained significant $(F(1, 5025) = 91.87, p < .001)$, with interpolation tasks still completed faster than concatenation tasks. To evaluate task accuracy, we assessed the association between participants' responses and the correct answers using Pearson's correlation. The correlation was positive and statistically significant, $r(5038) = .55$, $p < .001$, indicating that participants' answers aligned well with the correct outputs across tasks.

\begin{figure}
  \centering
  \includegraphics[width=0.80\linewidth,alt={Comparison of task completion duration across task complexity levels and conditions.}]{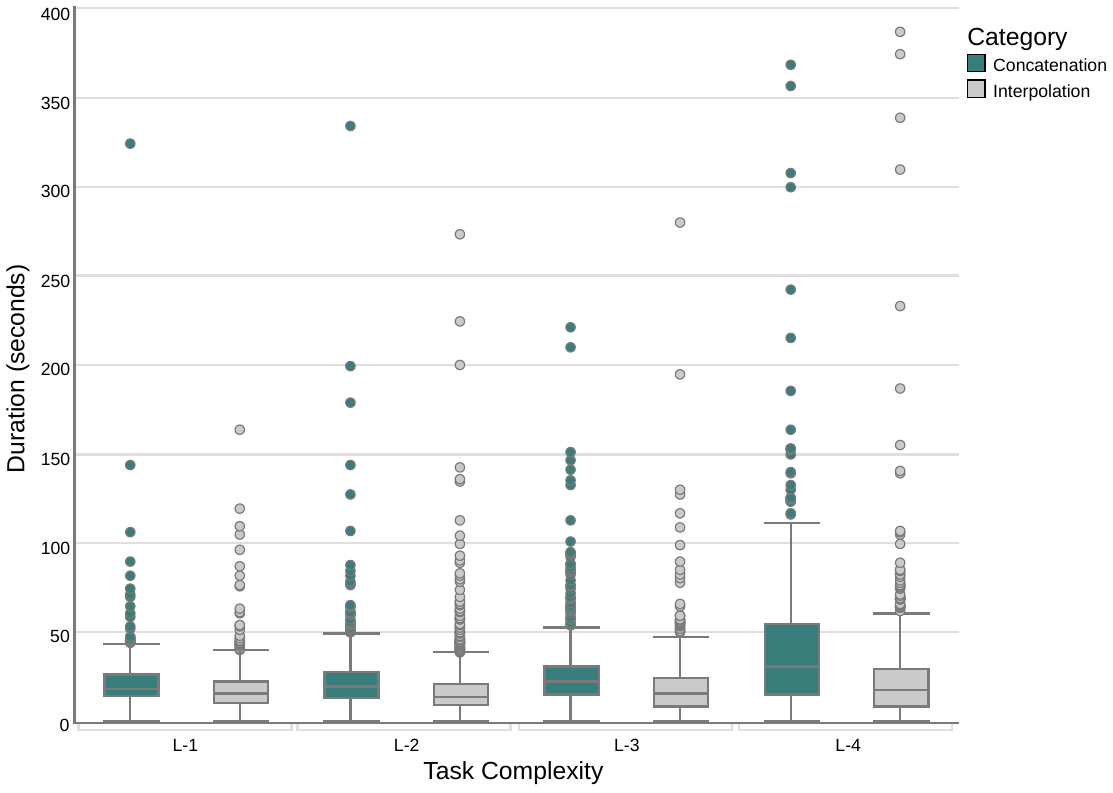}
  \caption{Distribution of task completion duration across task complexity levels (L--1 to L--4) for both concatenation and interpolation categories. Each boxplot represents the spread of completion times (in seconds), with outliers shown as individual points. As task complexity increased, completion times also tended to rise, particularly for concatenation tasks.}
  \Description{Comparison of Task Complexity and  Completion Time for Each Category}
  \label{Fig:TaskLevel-Duration}
\end{figure}

We were also interested in understanding the impact of task complexity on completion time, illustrated by a box plot in Figure \ref{Fig:TaskLevel-Duration} showing complexity levels from L--1 to L--4 (cf. Table~\ref{Tab:TaskComplexity}) against completion times. A One-Way repeated measures ANOVA was conducted on \textit{task complexity} across 4 independent samples. The result indicated that there was a statistically significant difference between the samples: \textit{task complexity}: $F(3, 5036) = 57.32, p < .001, \eta^2_G = .033$.

\begin{custombox}
{\faLightbulbO\hspace{0.01in} \textbf{Summary of RQ1 Results:}}
\textit{Embedding variables within string literals significantly affects program readability. Interpolation tasks were completed faster than concatenation tasks ($F(1, 5026) = 92.35$, $p < .001$, $\eta^2_G = .018$), with high accuracy maintained across both conditions ($r = .55$, $p < .001$; overall accuracy $\approx 90.6\%$), indicating reduced cognitive load under interpolation. Academic level showed a small but significant effect ($F(6, 5026) = 2.41$, $p = .025$, $\eta^2_G = .003$); post-hoc Tukey–Kramer test revealed only one significant pairwise difference (Junior vs.\ Senior, $p = .044$), and this effect was negligible in magnitude. Task complexity had a significant effect on completion time ($F(3, 5036) = 57.32$, $p < .001$, $\eta^2_G = .033$), with higher complexity yielding longer times particularly for concatenation. The primary advantage of interpolation was robust to substituting programming experience for academic level as a covariate ($F(1, 5025) = 91.87$, $p < .001$). These findings support $H_1$.}
\end{custombox}
 
\subsection{Results of Qualitative Analysis}
\vspace{0.5\baselineskip}
\begin{framed}
\underline{\textbf{RQ2:}} Relative to a concatenation baseline, how does literal string interpolation influence programmers' perceived readability, comprehension speed, debugging ease, and error propensity across tasks of varying complexity and levels of prior familiarity?
    \begin{itemize}
        \item[\textit{H\textsubscript{0}:}] \textit{There is no difference between interpolation and concatenation on all primary outcomes (readability, time-to-understand/cognitive load, debugging ease, error rate).}
        \item[\textit{H\textsubscript{1}:}] \textit{Interpolation outperforms concatenation on at least one primary outcome (higher readability, faster understanding/lower cognitive load, easier debugging, or fewer errors).}
     \end{itemize} 
    
\end{framed}

\noindent This section reports the findings from the analysis of the open-ended post survey questions. Out of 314 responses, we excluded 10 responses which contain null and random text responses which are not valuable to the analysis. The remaining 304 responses formed our study population for qualitative analysis. For each open-ended question, we independently sampled the data using a 95\% confidence level and a 5\% margin of error. Thematic coding was conducted using a structured framework based on our research questions. As aforementioned, we did not compute Cohen's Kappa due to the interpretive nature of the task, agreement was reached through coder discussion and convergence on final themes. Details of the coding procedure and reviewer roles are described in Section~\ref{sec:qual}. In the results, we provide concrete examples and notable excerpts from respondents. Themes are highlighted in \textbf{bold fonts}, conversations are presented in \textit{italics} and preceded by a person icon \faUser, highlighted excerpts are \textbf{emphasized in bold}, and longer messages are abbreviated with ellipses [...]. Additional examples and detailed summaries are available in the replication package~\cite{replication}, which includes all summaries and identified themes. To support cross-referencing with the replication package, each instance is labeled with a unique identifier: R-X (X being the instance number) of response. 

\subsubsection{Readability Reflection}\leavevmode\par 
\begin{framed}
\faQuestionCircle\hspace{0.02in} \textbf{Survey Question 1:}  Based on your experience with both string interpolation and string concatenation in the survey, which approach did you find more readable, and why? Please provide specific examples or reasons to support your viewpoint.
\end{framed}

\begin{figure}[htbp]
    \centering
    \includegraphics[width=\linewidth,alt={Figure summarizing participant responses about the readability of interpolation and concatenation.}]{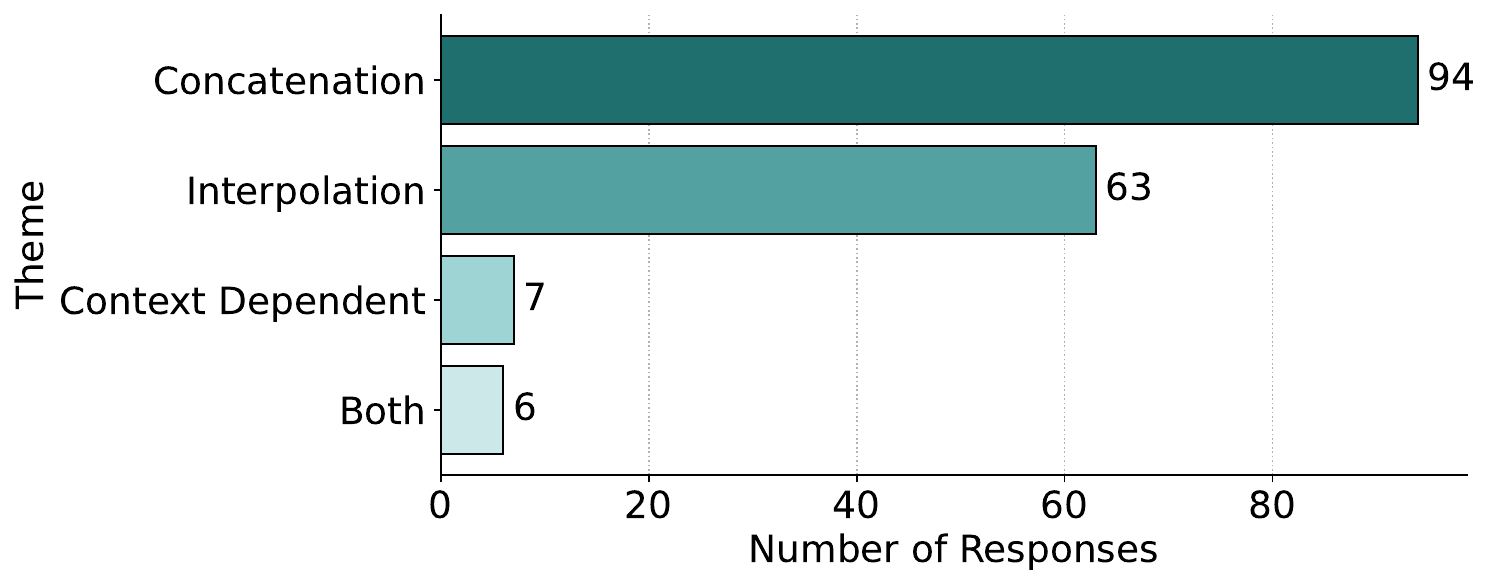}
    \caption{Distribution of readability preferences. Most participants favored string concatenation, followed by interpolation, while a smaller number expressed context-dependent or mixed preferences.}
    \Description{Readability reflection}
    \label{fig:readability-reflection}
\end{figure}

\noindent We analyzed 170 valid responses to examine how respondents perceive the readability of string concatenation and interpolation based on the sampling criteria described above. The analysis yielded four distinct preference groups, as shown in Figure~\ref{fig:readability-reflection}. Overall, 94 (55\%) respondents favored string \textbf{concatenation}, 63 (37\%) preferred \textbf{interpolation}, while smaller groups expressed \textbf{context-dependent} (7 respondents, 4\%) and 6 (4\%) respondents considered \textbf{both} constructs. These findings indicate that readability judgments are shaped by both syntactic familiarity and contextual complexity rather than a universally preferred approach.

Respondents who preferred concatenation often cited familiarity and habit as central reasons for their choice. The respondents accustomed to traditional operators (\texttt{+}) described concatenation as more intuitive and predictable, even if it introduced minor clutter. For instance, respondent R-142 explained: \faUser\hspace{0.03cm}\textit{``I found string concatenation more readable [...]. \underline{It feels straightforward} and I don't have to think  [...].''} This sentiment highlights how readability is often equated with cognitive ease and routine rather than structural simplicity. Conversely, participants favoring interpolation emphasized that embedding variables directly within text enhances clarity and reduces visual noise. They described interpolation as ``cleaner,''  ``more structured,'' and better aligned with modern programming conventions. For instance, R-113 stated: \faUser\hspace{0.03cm}\textit{``I found string interpolation \underline{more readable because it looks cleaner} and flows like a regular sentence. For example, f"Hello, \{name\}!" is easier to read than "Hello, " + name + "!" and [...].''} This group tended to associate readability with reduced syntactic friction and more immediate mapping between variables and output. 

A smaller subset adopted a context-dependent perspective, noting that readability varies with the complexity of the task. Concatenation was considered sufficient for simple or short outputs, while interpolation was preferred for longer strings with multiple variables. As R-102 reflected: \faUser\hspace{0.03cm}\textit{``\textbf{When the string was longer I had an easier time with interpolation but when they were shorter concatenation was easier} [...].''} Similarly, a few respondents considered both concatenation and interpolation to be equally readable, citing no clear preference. For example, respondent R-23 stated: \faUser\hspace{0.03cm}\textit{``They \textbf{both felt equally as readable} once I got the hang of it, [...]''}.

These findings suggest that readability perceptions are not governed by syntax alone but by how closely an approach aligns with developers' habits, code complexity, and tooling expectations. While concatenation benefits from familiarity and predictability, interpolation increasingly appeals in contexts demanding clarity and reduced cognitive load, a point we further elaborate in Section~\ref{sec:discussion} when discussing the implications of the findings.

\begin{custombox}
{\faLightbulbO\hspace{0.01in} \textbf{Summary of Readability Reflection Results:}}
\textit{The analysis shows a near balance between preferences for string concatenation (55\%) and interpolation (37\%), with some respondents favoring context-dependent use (4\%). Concatenation was valued for familiarity and habit, while interpolation was applauded for clarity, reduced clutter, and structural readability. Key sub-themes included general readability rationale, clarity of structure, and familiarity. 
}
\end{custombox}
\subsubsection{Comprehension and Debugging}\leavevmode\par 
\begin{framed}
\faQuestionCircle\hspace{0.02in} \textbf{Survey Question 2:} Did using string interpolation or concatenation affect your ability to quickly understand or debug code snippets? Please describe any challenges or advantages you experienced.
\end{framed}

\begin{figure}[htbp]
    \centering
    \includegraphics[width=\linewidth,alt={Figure summarizing participant responses about comprehension and debugging.}]{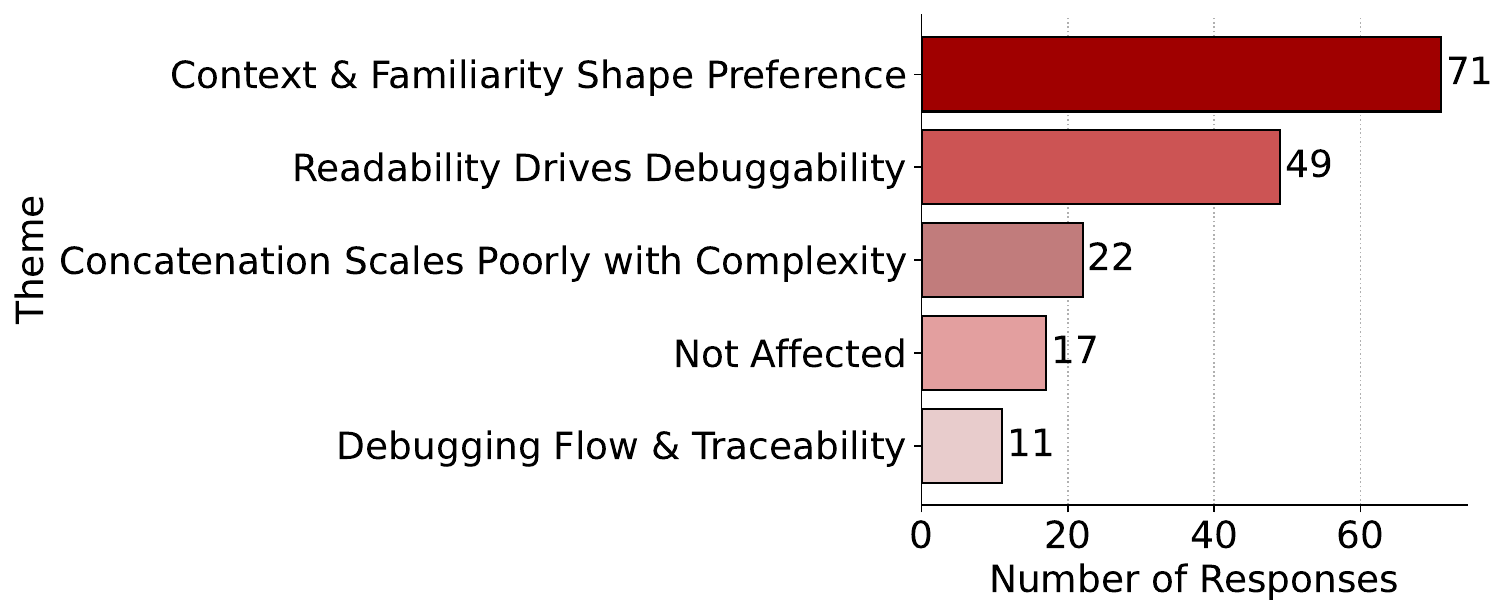}
    \caption{Distribution of themes related to comprehension and debugging. Most participants emphasized the influence of familiarity and readability, followed by complexity challenges, no perceived effect, and traceability benefits.}
    \Description{Comprehension and debugging reflection}
    \label{fig:rq2-themes}
\end{figure}

\noindent From the 170 sampled valid responses, our analysis identified five recurring themes shaping participants' perceptions of comprehension and debugging. These themes illustrate how familiarity, readability, and code complexity jointly influence developers' debugging efficiency. Figure~\ref{fig:rq2-themes} shows the distribution of responses across these themes.
We observed that the most prevalent theme is \textbf{context and familiarity shape preference}, accounting for 71 responses (42\%). This theme emphasizes how prior experience and habit strongly influenced perceived debugging ease. Respondents often found whichever method they were most accustomed to (typically string concatenation) easier to follow, even when not structurally optimal. For example, participant R-93 explained: \faUser\hspace{0.03cm}\textit{``string \textbf{interpolation examples took more time}, [...]. \textbf{Not used to it} so I think my brain kinda melted [...]...''}. This pattern highlights that debugging comfort often stems more from learned familiarity than from intrinsic code readability.

The second major theme, \textbf{readability drives debuggability}, was identified in 49 responses (29\%). Participants described string interpolation as improving debugging by reducing visual clutter and clarifying variable-to-text mappings. Readability was consistently linked to faster error detection and reduced mental effort. As respondent R-87 noted: \faUser\hspace{0.03cm}\textit{``Yes, I think it took me a bit longer to read a concatenated snippet. \textbf{Interpolation reduced the visual noise and made the mapping between variables and text clearer}, which made debugging easier.''} This group tended to associate interpolation with a more direct and cognitively efficient debugging process, particularly in multi-variable or complex code segments. A smaller but notable subset of 22 respondents (13\%) observed that \textbf{concatenation scales poorly with complexity}. While concatenation was considered manageable for short strings, participants found it error-prone and harder to maintain as the number of variables increased. Common issues included operator clutter, spacing errors, and difficulty tracing variable flow. One respondent (R-120) summarized this challenge succinctly: \faUser\hspace{0.03cm}\textit{``Yes, interpolation made debugging easier. \textbf{When using concatenation, I had to mentally piece together each variable and string fragment, which got confusing} [...]''}

Another 17 respondents (10\%) reported \textbf{no perceived effect}, indicating that the choice between interpolation and concatenation did not substantially influence their debugging efficiency. These participants attributed debugging performance more to factors such as overall code organization, naming consistency, and logical structure than to string syntax. For this group, readability and debugging were seen as broader design concerns rather than language-specific issues.

Finally, 11 respondents (6\%) emphasized \textbf{debugging flow and traceability}, suggesting that concatenation sometimes aided execution tracing in simple logging or output scenarios. However, these advantages diminished as projects grew in size and complexity. One participant explained: \faUser\hspace{0.03cm}\textit{``Concatenation sometimes makes it easier to trace execution flow in simple logs, but once the code gets more complex, I lose track quickly and prefer interpolation.''} This theme underscores that while concatenation may occasionally enhance traceability in isolated cases, interpolation generally supports clearer and more maintainable debugging.

\begin{custombox}
{\faLightbulbO\hspace{0.01in} \textbf{Summary of Comprehension \& Debugging Results:}}
\textit{The analysis of 170 responses revealed five major themes shaping developers' debugging experiences. Familiarity (42\%) and readability (29\%) were the dominant influences, followed by challenges with complexity (13\%), no perceived effect (10\%), and traceability advantages (6\%). Overall, while concatenation benefits from familiarity and simplicity in smaller contexts, interpolation offers superior readability and scalability, making it more effective for complex debugging tasks.}
\end{custombox}

\subsubsection{Preference Rationale}\leavevmode\par 
\begin{framed}
 \faQuestionCircle\hspace{0.02in} \textbf{Survey Question 3:} If you were tasked with choosing between string interpolation and string concatenation for future programming tasks, which would you prefer, and what specific factors influenced your choice?
\end{framed}

\begin{figure}[htbp]
  \centering
  \includegraphics[width=0.9\linewidth,alt={Figure summarizing participants reasons for preferring interpolation or concatenation.}]{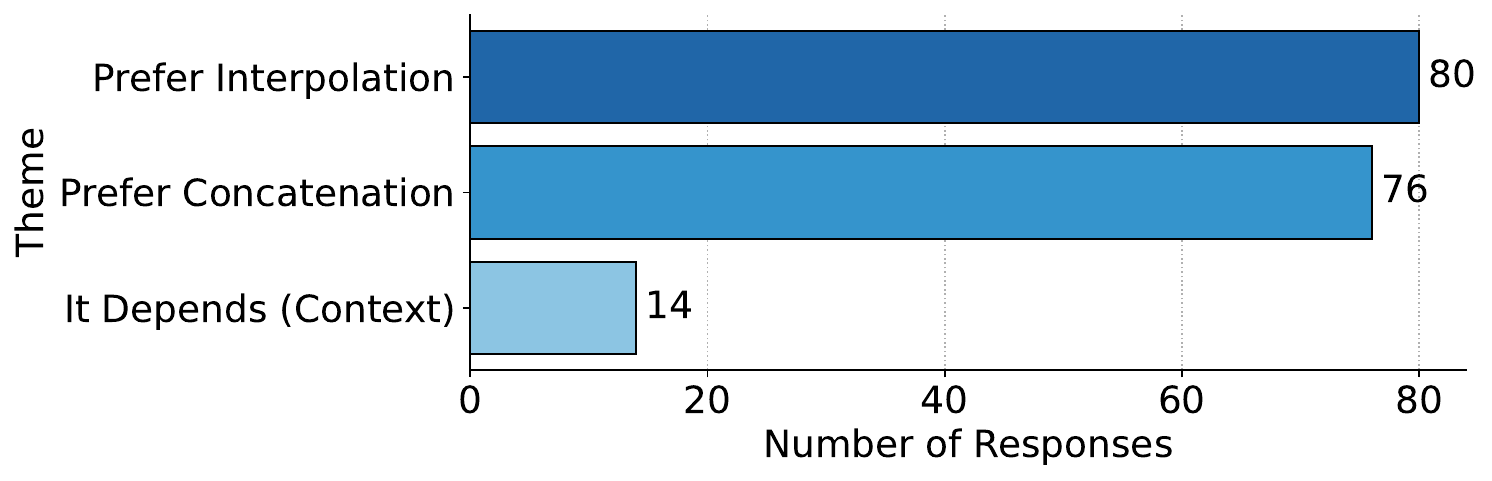}
  \caption{Distribution of themes related to preferences. Most participants preferred string interpolation, followed closely by string concatenation, while a smaller group indicated that their choice depends on context.}
  \Description{Preference rationale}
  \label{fig:rq3-pref-upd}
\end{figure}

\noindent From the 170 valid responses, preferences were nearly evenly divided: 80 respondents favored interpolation, 76 preferred concatenation, and 14 indicated that their choice depended on context. These findings highlight how developers weigh readability, familiarity, and situational context when selecting a preferred approach for future programming tasks (cf. Figure~\ref{fig:rq3-pref-upd}).

Respondents who preferred \textbf{string interpolation} (80 responses, 47\%) emphasized readability and clarity as the primary factors guiding their choice. Interpolation was consistently described as ``cleaner,'' ``less cluttered,'' and ``easier to understand at a glance.'' Several respondents associated it with modern language practices and noted that it minimizes the visual noise and operator overhead present in concatenated expressions. As one participant (R-212) explained: \faUser\hspace{0.03cm}\textit{``Interpolation. Strings are far easier to read and maintain when variables are directly embedded, without clutter from operators.''} This perspective frames interpolation as both a stylistic and ergonomic improvement that reduces debugging effort in complex, variable-rich code. In contrast, \textbf{string concatenation} remained the preferred method for 76 respondents (45\%), primarily due to familiarity, perceived simplicity, and predictability. Participants accustomed to traditional operators (\texttt{+}) expressed confidence in its directness and transparency, often framing interpolation as unnecessary abstraction. As one respondent (R-94) noted: \faUser\hspace{0.03cm}\textit{``[...], \textbf{string concatenation is easier to understand and feels straightforward for me}.''} For this group, readability was closely tied to procedural familiarity and control rather than syntactic minimalism. Their comments reflect how entrenched habits can outweigh structural advantages when evaluating perceived comprehension.

A smaller subset expressed a \textbf{context-dependent preference} (14 respondents, 8\%), emphasizing that the choice between concatenation and interpolation depends on the complexity of the task. Concatenation was considered sufficient for short, simple strings, while interpolation was viewed as more maintainable and readable in cases involving multiple variables. Respondent R-145 summarized this perspective: \faUser\hspace{0.03cm}\textit{``\textbf{String interpolation is easier to read, but concatenation works fine for shorter cases}. [...] The choice really depends on the complexity of the task.''} This view reinforces the notion that some developers calibrate their practices based on situational demands rather than adherence to a single preferred syntax.

\begin{custombox}
{\faLightbulbO\hspace{0.01in} \textbf{Summary of Preference Rationale Results:}}
\textit{The analysis of 170 responses revealed a near-even split in preferences: 47\% favored interpolation, 45\% preferred concatenation, and 8\% were context-dependent. Interpolation was valued for its readability, clarity, and alignment with modern syntax, whereas concatenation was preferred for its familiarity, simplicity, and predictability. Overall, developer preferences were shaped by task complexity and prior experience, suggesting that educators should emphasize comparative reasoning and that industry should promote consistent, readability-driven standards.}
\end{custombox}

\subsubsection{Learning Curve}\leavevmode\par 
\begin{framed}
\faQuestionCircle\hspace{0.02in} \textbf{Survey Question 4:} How did your prior familiarity or experience with string interpolation or concatenation impact your ease of understanding the provided code examples? Please elaborate on your experience.
\end{framed}

\noindent From the 170 valid responses, four themes emerged, showing how prior experience and code complexity together influenced comprehension. The most common theme, \textbf{Familiarity Boosted Understanding} (86 responses, 51\%), captured the effect of what many described as ``muscle memory.'' Participants said that prior exposure to a particular method, most often concatenation, made examples written in that style easier to follow. Familiarity built confidence and fluency, even when the alternative might have been structurally clearer. As one respondent (R-119) shared: \faUser\hspace{0.03cm}\textit{``\textbf{My familiarity with concatenation} made problems related to concatenation much easier and feel less stressful.''} This theme underscores how ingrained habits can speed up comprehension but may also reinforce less efficient approaches.

The second theme, \textbf{Context and Complexity Moderated Impact} (67 responses, 39\%), revealed that ease of understanding depended on the complexity of the example. Participants who were comfortable with both methods found short, single-variable strings equally clear, but as examples became longer or included more variables, interpolation was seen as more readable. Reduced operator clutter and clearer variable-to-text mapping were identified as key advantages. As R-48 explained: \faUser\hspace{0.03cm}\textit{``I have experience with both. \textbf{Concatenation was easy at first, but the longer the strings got, interpolation was easier} to figure out, just because of how lengthy the concatenations got.''}

\begin{figure}
  \centering
  \includegraphics[width=\linewidth,alt={Figure summarizing responses about the learning experience with the two string-construction styles.}]{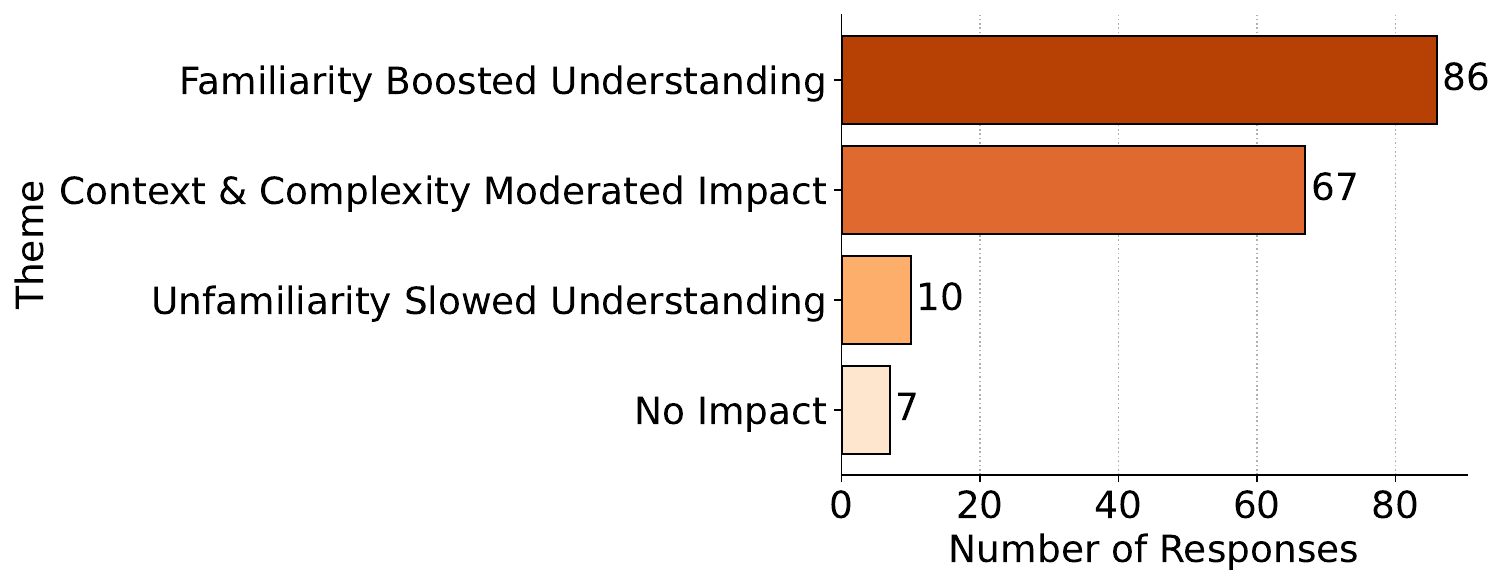}
  \caption{Distribution of themes related to the learning curve. Most respondents cited familiarity or context and complexity as key influences on understanding, while smaller groups mentioned unfamiliarity or no impact.}
  \Description{Learning curve}
  \label{fig:rq4-theme-counts}
\end{figure}

A smaller subset of participants (10 responses, 6\%) reflected the theme \textbf{Unfamiliarity Slowed Understanding}, describing moments of hesitation or added cognitive effort when they encountered a less familiar syntax, most often interpolation, after years of using concatenation. One participant (R-279) admitted: \faUser\hspace{0.03cm}\textit{``\textbf{I honestly forgot what string interpolation was} [...]. I used concatenation [...], and I still mostly use it now.''} This pattern suggests that limited exposure can reduce flexibility and make it harder to adapt to alternative syntactic styles. Finally, seven respondents (4\%) reported \textbf{No Impact}, indicating that their prior experience did not meaningfully affect comprehension. These participants noted that factors such as clear variable naming and overall code organization were more influential than the specific method used to form strings.
\begin{custombox}
{\faLightbulbO\hspace{0.01in} \textbf{Summary of Learning Curve Results:}}
\textit{The analysis of 170 responses identified four themes. Familiarity Boosted Understanding (51\%) showed that prior exposure accelerated comprehension, though sometimes reinforced less efficient habits. Context and Complexity Moderated Impact (39\%) indicated that both styles were clear for short strings, but interpolation was preferred for longer, variable-rich examples. Unfamiliarity Slowed Understanding (6\%) captured hesitation when encountering rarely used syntax, while No Impact (4\%) reflected neutral experiences. Overall, comprehension was shaped by both habit and structural complexity, with interpolation's readability advantages increasing as complexity rose. 
}
\end{custombox}

\subsubsection{Suggestions for Improvement}\leavevmode\par 
\begin{framed}
\faQuestionCircle\hspace{0.02in} \textbf{Survey Question 5:} Do you have any suggestions or recommendations on how either string interpolation or concatenation syntax could be improved to enhance readability and comprehension?
\end{framed}

\begin{figure}[htbp]
  \centering
  \includegraphics[width=\linewidth,alt={Figure summarizing participants suggestions for improving string syntax and tooling.}]{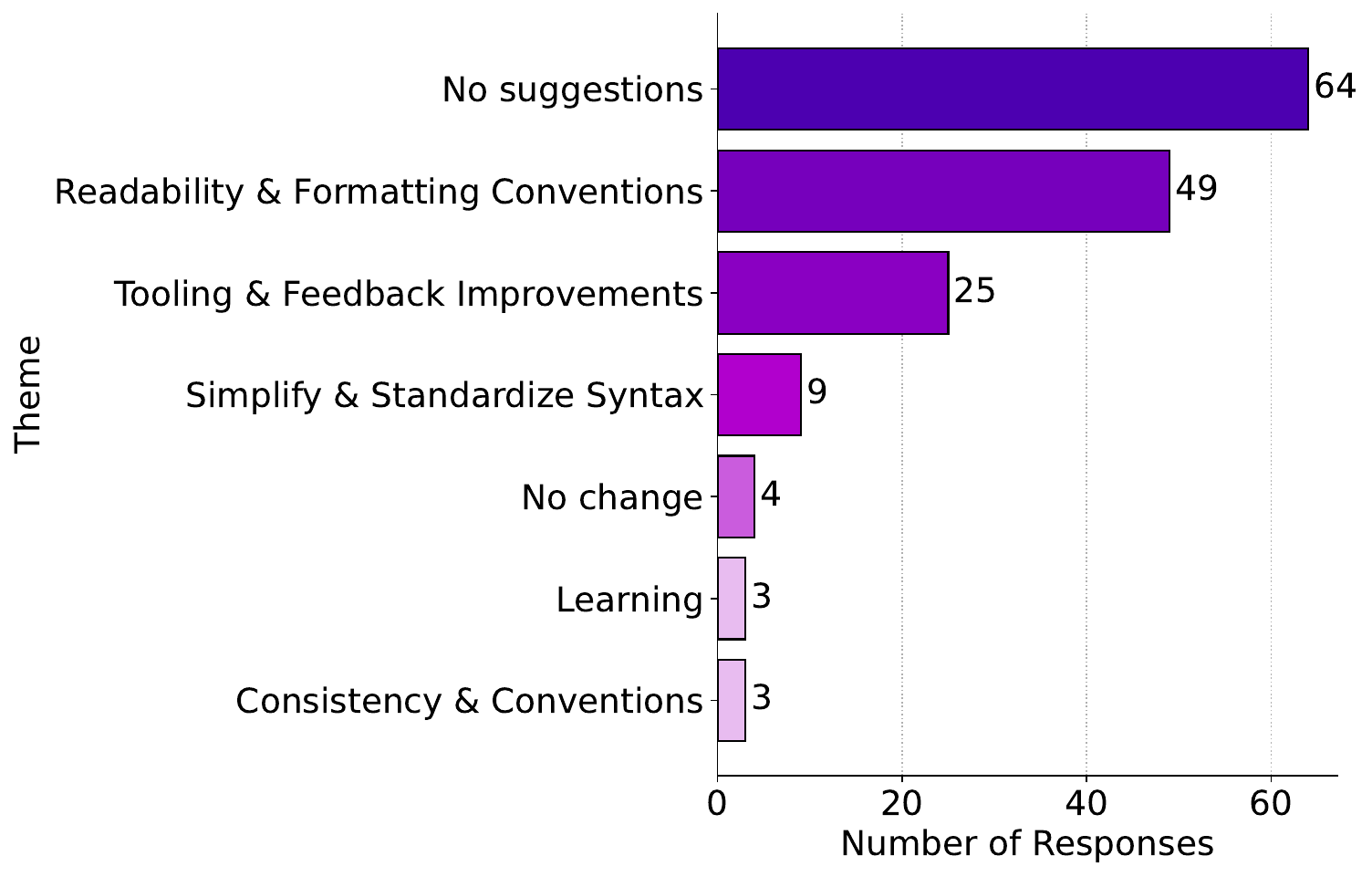}
  \caption{Distribution of suggestions for improvement. Most respondents offered no suggestions or emphasized readability and formatting, followed by tooling feedback, syntax simplification, and smaller clusters focusing on learning, consistency, and no change.}
  \Description{Improvement suggestions}
  \label{fig:rq5-impro}
\end{figure}

\noindent For this open-ended question, 264 responses were selected from the 304 total participants, as 37 individuals did not provide an answer. After excluding incomplete or invalid entries, a final sample of 157 valid responses was analyzed. The thematic analysis of these responses revealed seven recurring themes capturing both syntactic and environmental factors that influence code readability (cf. Figure~\ref{fig:rq5-impro}). The largest group amounting to 64 (38\%) responses offered \textbf{no suggestions}, expressing overall satisfaction with existing syntax or viewing readability as a matter of personal style and familiarity rather than a design issue. The second most common theme was \textbf{readability and formatting conventions} consisting of 49 (29\%) responses. They emphasized the need for clearer visual separation between literals and variables, reduced punctuation and spacing clutter, and more explicit or named placeholders to improve variable binding clarity in longer or more complex strings. For instance, respondent R-18 stated: \faUser\hspace{0.03cm}\textit{``[...] concatenation could be improved by \textbf{reducing the number of plus signs} [...] interpolation could be clearer with \textbf{more distinct formatting to separate variables}... [...].''}

A substantial number of participants also recommended \textbf{tooling and feedback improvements} (25 responses, 15\%), focusing on enhanced IDE/editor support, auto-formatting, live previews of evaluated strings, and linter rules that catch missing placeholders or inconsistent syntax. For example participant R-172 provided a comprehensive suggestion; 
\faUser\hspace{0.03cm}\textit{``For interpolation, [...]. One improvement could be \textbf{built-in formatting rules inside the \{ \} for better alignment or padding} [...]. Also, tools that \textbf{highlight interpolated variables} [...]. For concatenation, \textbf{perhaps simplifying how non-string types are automatically} handled could reduce the need for str() calls, [...]. Alternatively, providing \textbf{warnings or hints when mixing types during concatenation} could guide less experienced users''.}

These suggestions highlight the perceived value of tooling as a readability enabler rather than purely a development aid. A smaller group (9 responses, 5\%) advocated for \textbf{simpler, standardized syntax}, such as lighter-weight delimiters, less ambiguous braces, and convergence on common string-handling conventions across languages. Additional minor themes included calls for improved \textbf{learning resources} (4 responses, 2\%), greater \textbf{consistency and conventions} across frameworks (3 responses, 2\%), and affirmations of \textbf{no change necessary} (3 responses, 2\%). Together, these responses demonstrate that while many developers are content with current language features, a significant subset see opportunities for improvement through enhanced readability practices and stronger tool integration.

Overall, these perspectives suggest that readability is a shared responsibility between syntax design and the surrounding tooling ecosystem. When variables are visually distinct and semantically clear, comprehension improves; when punctuation and operators accumulate, it deteriorates. Developers thus envision improvement not only through new syntax but through environments that actively enforce and visualize clarity.

\begin{custombox}
{\faLightbulbO\hspace{0.01in} \textbf{Summary of Suggestions for Improvement Results:}}
\textit{The analysis revealed seven themes reflecting developers' perspectives on improving string readability. Most respondents (38\%) offered no suggestions, while 29\% emphasized readability and formatting conventions—calling for clearer separation between literals and variables and less punctuation clutter. 15\% focused on tooling enhancements such as IDE highlighting and linting, and 5\% proposed simpler, standardized syntax. Smaller groups stressed better learning resources 2\%, consistency across languages ~2\%, or saw no need for change 2\%. Overall, participants viewed readability as a joint product of syntax and tooling, advocating for conventions, visual feedback, and cross-language consistency to enhance comprehension and maintainability.}
\end{custombox}

\section{Discussion}
\label{sec:discussion}
This section synthesizes the quantitative and qualitative results to explain how string interpolation and concatenation influence programmers' comprehension and learning. The integrated evidence shows that interpolation enhances efficiency and readability, while familiarity biases shape learners' syntax preferences. The following subsections interpret these findings in relation to the study's hypotheses, connect them to cognitive-load and readability theories, and outline implications for programming education and practice.

\subsection{Support of Hypotheses and Interpretation}

The first research question (RQ1) examined whether embedding variables, expressions, and functions within string literals influences program readability and comprehension. The quantitative results show that syntax choice meaningfully affects how learners engage with and reason about code. Tasks using string concatenation consistently required more time to complete, reflecting greater cognitive effort in parsing fragmented syntax. As task complexity increased, learners took longer to reason through concatenation-based examples (cf.\ Figure~\ref{Fig:TaskLevel-Duration}), suggesting that concatenation imposes additional demands as learners must manage a growing number of tokens, operators, and delimiters. In contrast, interpolation supported more immediate pattern recognition by presenting related elements in a cohesive structure. Together, these findings support the alternative hypothesis (\textit{H1}), indicating that syntactic design can significantly influence comprehension efficiency. Consistent with this result, analyses for RQ2 also supported rejection of the null hypothesis (\textit{H0}), showing that interpolation improved comprehension speed without reducing accuracy.

Across task categories, completion-time distributions exhibited overlapping variability (cf.\ Figure~\ref{Fig:BoxPlotOverall}), yet participants consistently completed interpolation tasks faster than concatenation tasks. This pattern suggests that interpolation reduces syntactic overhead by minimizing the need to track quotation marks, operators, and spacing, thereby streamlining cognitive processing during task execution. Although completion times varied across academic levels (cf.\ Figure~\ref{Fig:YearInCollege-Duration}), the advantage of interpolation remained consistent, indicating that its readability benefits generalize across experience levels. Moreover, the moderate positive association between participants' responses and correct answers demonstrates that learners generally understood the tasks well, maintaining high accuracy across conditions.

Task complexity also played a significant role in shaping performance. As shown in Figure~\ref{Fig:TaskLevel-Duration}, completion time increased steadily from lower-complexity tasks (L--1) to higher-complexity tasks (L--4). This pattern aligns with established findings in programming cognition, which show that increased structural complexity slows comprehension and raises processing demands~\cite{pennington1987comprehension,storey1999cognitive,schraw1998metacognition}. These results suggest that learners benefit when syntax reduces extraneous cognitive load and supports perceptual grouping.

From a learning-science perspective, the findings can be interpreted through theories of schema formation, perceptual chunking, and working memory~\cite{SchemaTheory,Chunking,Baddeley2000WorkingMemory}. They also line up with empirically grounded readability models~\cite{eye-tracking}. Buse and Weimer's~\cite{Buse2010} metric weights operator and punctuation density heavily, predicting that concatenation, dense in `+' operators and quotation marks, should be rated as less readable than the more uniform surface form of interpolation. Scalabrino et al.'s~\cite{Scalabrino2016} textual-feature model points the same way: it emphasizes the cognitive value of keeping semantically related tokens close to each other, and interpolation places each variable next to its surrounding text where concatenation breaks that context apart across operators. The faster completion times we observed for interpolation, particularly at higher complexity, are what both models would predict. Developers familiar with concatenation may still process it efficiently through established schemas, despite its syntactic density, but interpolation's compact form supports chunking by grouping variables and literals into a single unit, lowering working-memory demands during comprehension and debugging. This interpretation is consistent with prior work showing that structural clarity and naming conventions affect readability and developer performance as complexity grows~\cite{Lawrie2006,Binkley2013}. For educators, the takeaway is that syntactic choices are not neutral: they shape the mental models and comprehension strategies students carry across languages.

\subsection{Similarity of Results}
These findings are consistent with prior research on programming language readability and comprehension, which similarly emphasizes the role of syntactic complexity in shaping cognitive effort. Numerous studies have examined the impact of programming language syntax on readability and comprehension, exploring a variety of topics including lambda expressions~\cite{hanenberg2022two,10.1145/3133909}, polyglot programming~\cite{10.1145/3368089.3409701}, regular expressions \cite{8091188}, and various programming language features~\cite{zid2024study,10.1145/3291279.3339421}, among others. There is a general consensus that code complexity hinders readability and comprehension. For example, the study cited in~\citet{10.1145/3341525.3387384} demonstrates that reading short and specific compiler error messages leads to faster debugging times and higher usefulness scores among participants. This aligns with our observations shown in Figure~\ref{Fig:TaskLevel-Duration}, where task complexity appears to influence completion time.

Our findings regarding the difference in task completion time between freshmen and participants from upper college years also align with the observation that novice developers often make more programming mistakes, as they are still learning various programming constructs~\cite{10.1145/3335814}. This emphasizes the learning curve associated with gaining proficiency in programming, which can significantly affect performance on programming tasks.

\subsection{Threats to Validity}

 First, task completion time and correctness provide objective measures of program comprehension, but they do not capture all cognitive processes involved in understanding code. To complement these measures, we analyzed participants' qualitative reflections, allowing us to interpret quantitative performance alongside their reported reasoning. We report both the proportion of correct responses and the correlation between participants' answers and the expected answers using Pearson's~$r$ (Section~\ref{sec:statistics}, Table~\ref{Tab:ExtendedTaskStats}), providing complementary perspectives on participant performance.
 
Second, we used a language-neutral notation to compare concatenation and interpolation without introducing bias from participants' familiarity with a particular programming language. This abstraction may have required some participants to spend effort understanding the notation rather than the underlying construct. To reduce this risk, participants completed practice tasks before the study, and ambiguous items identified during the pilot were revised or removed. A related concern is that the task instructions themselves may anchor how participants judge the code: \citet{Wyrich2022anchoring} show that the wording of a task description can shift code-understandability evaluations. Because every task in our study used the same neutral instruction template and asked only for the resulting string, this framing was held constant across the concatenation and interpolation conditions, so any anchoring effect applies equally to both and does not bias the within-subjects
comparison.

Third, our results apply to the conceptual distinction between concatenation and interpolation rather than to the syntax of any specific programming language. The magnitude of the observed effects may differ in studies using concrete language syntax (e.g., Python f-strings versus \texttt{+} concatenation). Generalizability is also limited by the use of 16 short comprehension tasks and a participant pool drawn primarily from one university, supplemented by volunteers recruited through LinkedIn. Finally, academic year is an imperfect proxy for programming experience. To assess whether this choice influenced our conclusions, we repeated the primary analysis using self-reported years of programming experience as a covariate (Section~\ref{sec:statistics}); the effect of task category remained unchanged.

Lastly, the qualitative analysis was assisted by an LLM but remained under human supervision throughout. All LLM-generated codes were reviewed by the authors, 30\% of the responses were independently double-coded, and disagreements were resolved through discussion. Because LLM-assisted qualitative coding is still an emerging practice with limited evidence on reliability~\cite{BanoMuneera}, we treat the resulting themes as interpretive rather than definitive and describe the safeguards used to improve trustworthiness in Section~\ref{sec:trustworthiness}.

\subsection{Trustworthiness of Qualitative Findings}\label{sec:trustworthiness}

For the qualitative component, we followed established principles of credibility, dependability, confirmability, and transferability~\cite{lincoln1985naturalistic,runeson2009guidelines}. Credibility was supported through iterative coding and researcher triangulation, where two coders independently analyzed 30\% of the data and resolved differences through discussion. Dependability was achieved by maintaining a clear audit trail, which included LLM-generated draft codes, refinement logs, and the final validated codebook. All of these materials are included in the replication package to support transparency and reuse. Confirmability was strengthened by grounding each theme in direct participant quotations and documenting the reasoning behind every decision to merge or refine codes. Transferability was enhanced by providing detailed descriptions of the tasks and participant demographics, allowing other researchers to reproduce the study in similar educational or industrial contexts.

Overall, our goal was to maintain both rigor and transparency through a mixed-method randomized controlled trial that integrated quantitative performance data with qualitative reasoning. The main remaining threats relate to ecological validity, specifically how well the simplified code snippets represent real-world debugging situations, and to potential bias in participants' self-reported reflections. Even with these limitations, the strong alignment between the quantitative results and the qualitative themes increases confidence in the robustness and interpretability of our findings.

\subsection{Implications}
\label{sec:implications}

The implications of this study extend beyond empirical performance differences to questions of how programming constructs are taught, learned, and adopted in practice. We frame these implications in terms of how evidence-based findings can inform instructional design and professional conventions, without overgeneralizing beyond the scope of the data.

\subsubsection{Implications for Programming Education}

Our findings suggest that string interpolation offers readability advantages over concatenation, particularly as code becomes more complex, but that learners' preferences are strongly shaped by early exposure and familiarity. From an educational perspective, this highlights the importance of how and when syntactic alternatives are introduced in programming curricula. Students who primarily encounter concatenation early may internalize it as the default approach, even when more readable abstractions are available.

Rather than focusing on syntax in isolation, instructors may benefit from explicitly contrasting alternative constructs and making readability an explicit learning objective. Prior work in computing education and metacognition emphasizes that students develop stronger programming judgment when they are prompted to reflect on their choices and their consequences for comprehension and maintenance~\cite{schraw1998metacognition, loksa2022programming}. Side-by-side examples, short refactoring exercises, and guided debugging tasks that compare concatenation and interpolation can help students recognize when higher-level abstractions reduce cognitive effort. Importantly, these activities need not be extensive; even brief, targeted comparisons can support learners in developing transferable reasoning about code clarity rather than relying solely on habit.

\subsubsection{Implications for Professional Practice and Tooling}

For professional software development, the results underscore that readability is not merely a stylistic preference but a factor that influences comprehension speed, debugging effort, and long-term maintainability. Teams may benefit from adopting clear, context-sensitive conventions that favor interpolation for complex or variable-rich strings while allowing concatenation in simpler cases. Such conventions can be reinforced through linters, style guides, and automated refactoring tools.

At the tooling level, participants' qualitative feedback highlights opportunities for IDEs and language ecosystems to better support readable string construction through placeholder highlighting, inline previews, and safe transformations between concatenation and interpolation. Gradual adoption strategies, supported by on-boarding materials and tooling assistance, may help mitigate resistance rooted in prior familiarity. 

In general, these implications suggest that improving readability and comprehension in string handling requires coordinated attention across education, industry, and language design. Educators can help developers understand readability trade-offs, while organizations and tool creators can institutionalize those lessons by embedding clarity, consistency, and cognitive efficiency into everyday software development practice. This is increasingly relevant as AI assistants and coding agents generate code that developers must read, adapt, and review before integration~\cite{ogenrwot2026patchtrack,ogenrwot:2024ase,ogenrwot2026how}.

\section{Conclusion}\label{sec:conclusion}

This study examined how string interpolation and string concatenation affect program readability, comprehension, and debugging efficiency using a mixed-method approach. Drawing on both performance data and qualitative feedback from 314 participants, we found that string interpolation generally supports faster comprehension than concatenation without compromising accuracy. These advantages held across different levels of programming experience and became more noticeable as tasks grew more complex, suggesting that interpolation is especially well suited for variable-rich or syntactically dense code.

The qualitative findings help explain these performance differences. Many participants described interpolation as cleaner and easier to follow, while others preferred concatenation largely out of habit and prior exposure. Participants also pointed to the role of tooling and conventions, such as formatting, variable highlighting, and IDE support, in shaping how readable code feels in practice. Overall, our results suggest that string interpolation offers clear readability benefits, but that its effectiveness depends not only on syntax, but also on experience and development context. 

\section{Future work}
Future work should examine how different string interpolation syntaxes affect readability and comprehension, helping identify designs that best support understanding and ease of use. In addition, studying the errors developers make when using interpolation, along with their debugging strategies and cognitive processes, for example through eye-tracking~\cite{eye-tracking} or cognitive-load measures, could inform clearer guidelines and better tool support for real-world programming.

\section*{DATA AVAILABILITY}

The dataset, codebook, anonymized qualitative responses, analysis scripts, and replication materials supporting the findings of this study are available in the accompanying replication package hosted on Zenodo~\cite{replication}. 

\section*{ACKNOWLEDGMENTS}
This work is supported by the National Science Foundation under Grant Numbers NSF\# 2542438. The authors thank the study participants for their time and thoughtful contributions. We also acknowledge the reviewers and colleagues from the University of Nevada Las Vegas, who provided valuable feedback during the development of this work.

\bibliographystyle{ACM-Reference-Format}
\bibliography{main.bib}

\newpage
\appendix
\section{Interfaces of the custom-built data collection tool used in the experiment.}
\begin{figure}[htbp]
    \centering
    \includegraphics[width=0.8\linewidth,alt={Screenshot of the custom-built data collection platform used for the study.}]{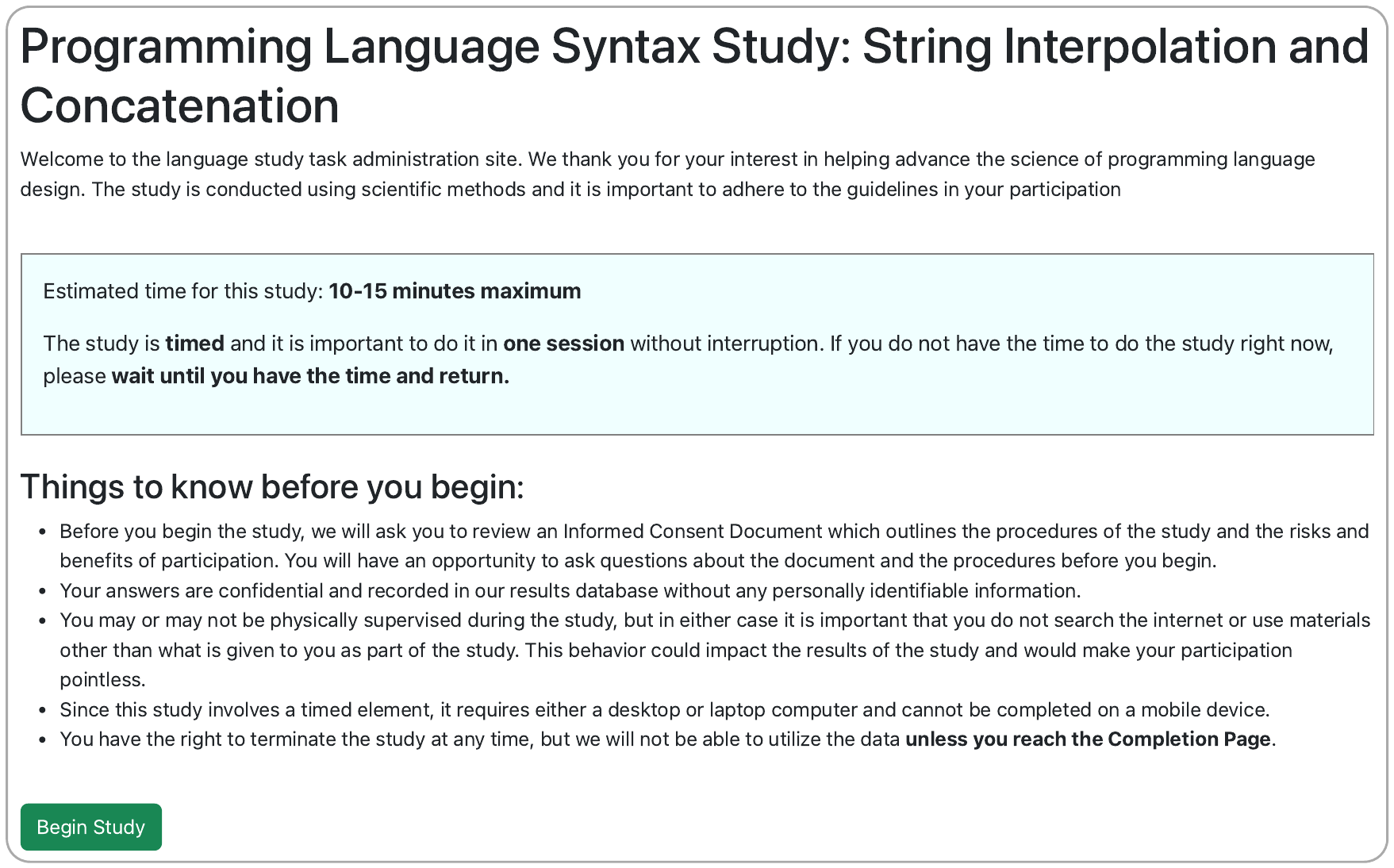}
    \caption{Interface of the custom-built data collection tool used in the experiment. The interface welcomes participants, provides instructions about the study's purpose and ethical considerations, and outlines participation requirements before proceeding to the demographic and practice task phases. 
    }
    \Description{Interface of the custom-built data collection tool used in the experiment.}
    \label{Fig:DataCollectionTool}
\end{figure}

\begin{figure}[htbp]
    \centering
    \includegraphics[width=0.65\linewidth,alt={Screenshot of the demographic survey interface used in the experiment.}]{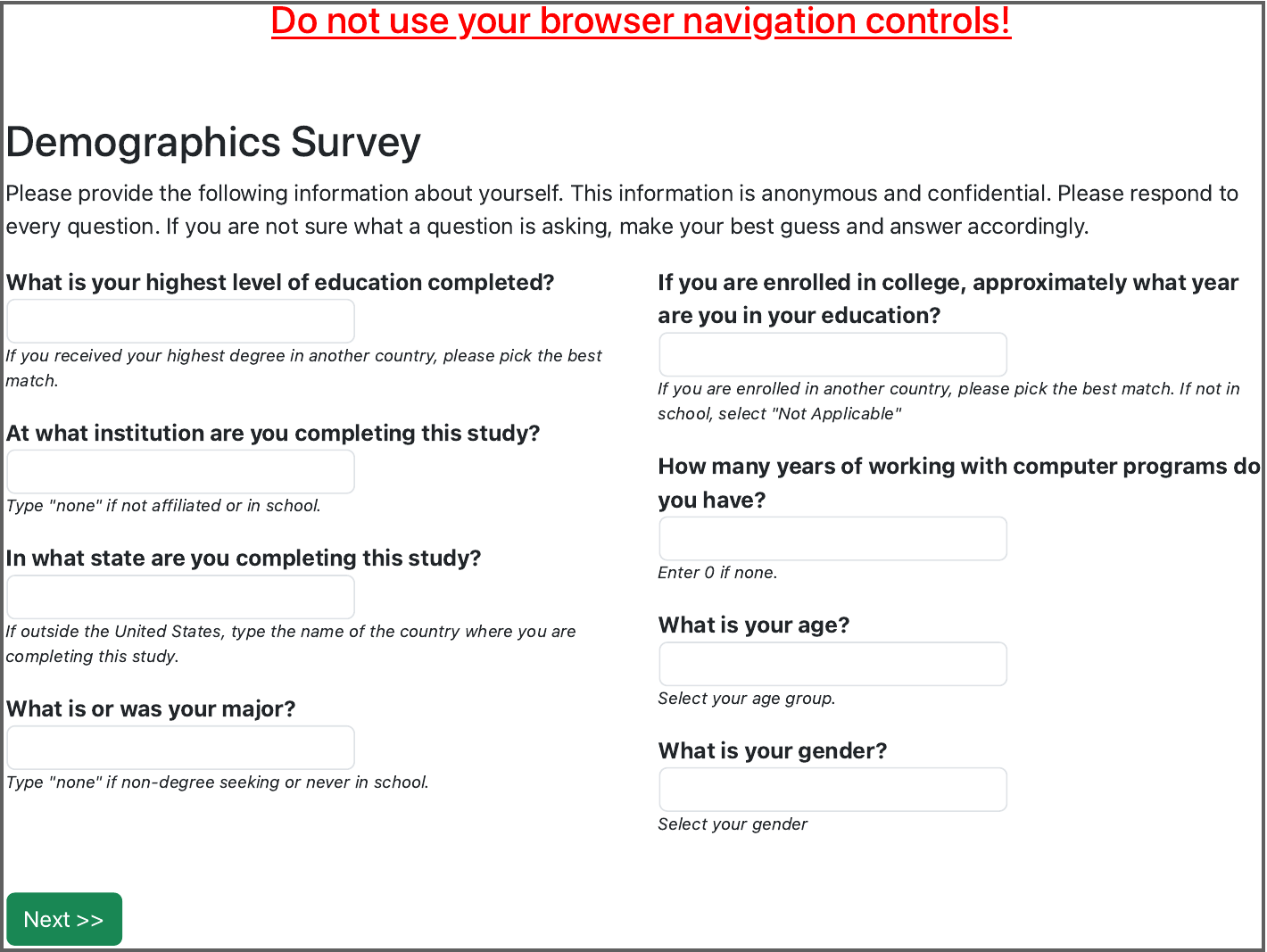}
    \caption{Screenshot of the demographic survey interface used in the study. Participants provided background information including education level, institutional affiliation, geographic location, major, academic year, programming experience, age, and gender. 
    }
    \Description{Screenshot of the demographic survey interface used in the study.}
    \label{Fig:Demographics}
\end{figure}
\section{Example of Codebook Used for Qualitative Analysis.}
\begin{landscape}
\begin{table}
\centering
\caption{Sample of the Readability Codebook Used for Qualitative Analysis.}
\label{tab:codebook-sample}
\footnotesize
\renewcommand{\arraystretch}{1.2}
\setlength{\tabcolsep}{4pt}

\tagpdfsetup{table/header-rows={1}}
\begin{tabularx}{\linewidth}{@{} l l X X X X @{}}
\toprule
\textbf{Theme} & \textbf{Sub-theme} & \textbf{Definition} & \textbf{When to Use} & \textbf{When Not to Use} & \textbf{Example Quote} \\
\midrule
Interpolation & Less boilerplate & Fewer tokens/joins; shorter, more concise construction reduces visual noise. & Mentions fewer concatenation operators or shorter code. & If terseness is not referenced. & ``Interpolation made it shorter to write and read.'' \\

Interpolation & Explicit pieces & Piecewise clarity; components are visibly separated. & Emphasizes clarity of separate parts or `+' between literals. & If the point is speed/cognitive effort without separation. & ``I found that it was more clear and I didn't have to think as hard.'' \\
Both/Context Dependent & Faster at a glance & Readable at first glance without detailed parsing. & Mentions quick or at-a-glance readability. & If timing refers to debugging, not reading. & ``\{a\}\{b\}\{c\} is much faster to understand than a+b+c.'' \\
None & Too verbose & Perceived wordiness or excessive length that hinders readability. & Calls out verbosity or clutter explicitly. & If brevity/verbosity isn't discussed. & ``The sentence felt long—especially because some of it was misspelled so it didn't match.'' \\
Both/Context Dependent & Fewer clutter & Reduced visual noise compared with multiple + operators and quotes. & Mentions `cleaner' look or fewer symbols. & If symbol count isn't discussed. & ``Interpolation avoids a bunch of + operators and quotes, especially in longer strings.'' \\
Both/Context Dependent & Clearer syntax & Simpler or more recognizable placeholder markers. & Refers to simpler braces or markers as clearer. & If clarity refers only to IDE coloring. & ``With the plus symbol, I would often find myself confusing it with addition.'' \\
Both/Context Dependent & Natural flow & Sentence-like flow that matches how the output is read. & Mentions natural reading order. & If flow/read order isn't mentioned. & ``Placing variables inline made it easier to visualize the sentence.'' \\
Concatenation & Clearer syntax & Simpler or more recognizable placeholder markers in some contexts. & References explicit separation of parts. & If clarity refers only to colorization. & ``It's clearer to see the separation using the + symbol and spaces.'' \\
Concatenation & Faster at a glance & Readable at first glance without mental expansion. & Mentions quick understanding. & If comment is about debugging time only. & ``I could understand it faster with concatenation.'' \\
Concatenation & Predictable evaluation & Left-to-right evaluation is explicit. & Mentions control/ordering benefits. & If ordering isn't addressed. & ``I know exactly when each part is added.'' \\
Interpolation & Fewer mistakes & Reduces quoting/parentheses errors. & Mentions fewer quote/paren mistakes. & If error proneness isn't discussed. & ``Less likely to mess up quotes with interpolation.'' \\

\bottomrule
\end{tabularx}
\end{table}
\end{landscape}

\section{Survey Programming Tasks on String Concatenation and Interpolation}\label{appendix:taskslist}

This appendix presents the complete list of programming tasks used in the survey to assess participants' comprehension, readability, and accuracy when working with string concatenation and string interpolation. The tasks were balanced across both string-handling categories and varied in complexity (Levels L--1 to L--4) to capture differences in performance and reasoning across a range of difficulty levels. Following the programming tasks, the post-survey section included five open-ended questions designed to elicit participants' qualitative reflections on readability, comprehension, debugging, learning experience, and improvement suggestions.

\begin{llmprompt}[Quantitative Survey Questions]{prompt:survey_quant_questions}
\footnotesize
\begin{enumerate}

\item \textbf{Task 1 of 16: The ``+'' symbol represents concatenation. Press 1, 2, 3 or 4 to select the correct concatenation. }\\
\begin{verbatim}
"John Doe, " + "3" + "1"
\end{verbatim}
\emph{Options:}
\begin{enumerate}
  \item \texttt{``JohnDoe, 31''}
  \item \texttt{``John Doe, 31''}
  \item \texttt{``Doe ''}
  \item \texttt{``John There''}
\end{enumerate}

\item \textbf{Task 2 of 16: The ``\{ \}'' symbol represents interpolation. Press 1, 2, 3 or 4 to select the correct interpolation}
\begin{verbatim}
"{5}{6}, John Doe"
\end{verbatim}
\emph{Options:}
\begin{enumerate}
  \item ``\verb|56, John Doe|''
  \item ``\verb|John Doe, 56|''
  \item ``\verb|56, Doe John|''
  \item ``\verb|Doe John There|''
\end{enumerate}

\item \textbf{Task 3 of 16: The ``+'' symbol represents concatenation. Press 1, 2, 3 or 4 to select the correct concatenation.}\\
\begin{verbatim}
a = "123 Main St, "
b = "Springfield, "
c = "OH"

a + (b + c)
\end{verbatim}
\emph{Options:}
\begin{enumerate}
  \item ``\verb|123 Main St, (Springfield, OH)|''
  \item ``\verb|123 Main St,Springfield,OH|''
  \item ``\verb|123 Main St,(Springfield,OH)|''
  \item ``\verb|123 Main St, Springfield, OH|''
\end{enumerate}

\item \textbf{Task 4 of 16: The ``\{ \}'' symbol represents interpolation. Press 1, 2, 3 or 4 to select the correct interpolation.}\\
\begin{verbatim}
a = "123 Main St, "
b = "Springfield, "
c = "OH"

"{b} ({a} {c})"
\end{verbatim}
\emph{Options:}
\begin{enumerate}
  \item ``\verb|Springfield,(123 Main St),OH|''
  \item ``\verb|Springfield, (123 Main St, OH)|''
  \item ``\verb|123 Main St, (Springfield, OH)|''
  \item ``\verb|(123 Main St,Springfield),OH|''
\end{enumerate}

\item \textbf{Task 5 of 16: The ``+'' symbol represents concatenation. Press 1, 2, 3 or 4 to select the correct concatenation.}\\
\begin{verbatim}
name = "Daniel" 
score = 50
total = 100

"Hello " + name + ", your score is " + score + " out of " + total + "."
\end{verbatim}
\emph{Options:}
\begin{enumerate}
  \item ``\verb|Hello name, your score is score out of total.|''
  \item ``\verb|Hello Daniel, your 50 is score out of total.|''
  \item ``\verb|Hello Daniel, your score is 50 out of 100.|''
  \item ``\verb|Hello Daniel, your score is 100 out of 50.|''
\end{enumerate}

\item \textbf{Task 6 of 16: The ``\{ \}'' symbol represents interpolation. Press 1, 2, 3 or 4 to select the correct interpolation.}\\
\begin{verbatim}
name = "Daniel" 
score = 50
total = 100

"{name}, out of {total}, you scored {score}."
\end{verbatim}
\emph{Options:}
\begin{enumerate}
  \item ``\verb|Hello Daniel, your score is 50 out of 100.|''
  \item ``\verb|Daniel, you scored 50 out of 100.|''
  \item ``\verb|Daniel, out of 100, you scored 50|''
  \item ``\verb|Daniel, out of 100, you scored 100|''
\end{enumerate}

\item \textbf{Task 7 of 16: The ``+'' symbol represents concatenation. Press 1, 2, 3 or 4 to select the correct concatenation.}\\
\begin{verbatim}
"World, " + 1 + 2 + 3 + (1 + 1 + 1 + 1) + ("5" + (6 + 7))
\end{verbatim}
\emph{Options:}
\begin{enumerate}
  \item ``\verb|World, 123("5"(67))|''
  \item ``\verb|World, 123("5"))|''
  \item ``\verb|123(1111)("5"(67)), World|''
  \item ``\verb|World, 123(1111)("5"(67))|''
\end{enumerate}

\item \textbf{Task 8 of 16: The ``\{ \}'' symbol represents interpolation. Press 1, 2, 3 or 4 to select the correct interpolation.}\\
\begin{verbatim}
"{1} {2} {3} {({9} {9} {9} {1})} World { ({"5"} {(67)} )}"
\end{verbatim}
\emph{Options:}
\begin{enumerate}
  \item ``\verb|1 2 3 (9 9 9 1) World ("5" (67) )|''
  \item ``\verb|123(9991)World("5"(67))|''
  \item ``\verb|1 2 3 ({9} {9} {9} 1) World ("5" (67) )|''
  \item ``\verb|1 2 3 9 9 9 1 World "5"67|''
\end{enumerate}

\item \textbf{Task 9 of 16: The ``+'' symbol represents concatenation. Press 1, 2, 3 or 4 to select the correct concatenation.}\\
\begin{verbatim}
a = 5
b = 6
c = 4
d = 10

"a = " + a + " d = " + d  + " c = " +  c + " b = " + b
\end{verbatim}
\emph{Options:}
\begin{enumerate}
  \item ``\verb|a = 5 b = 6 c = 4 d = 10|''
  \item ``\verb|a = 5 d = 6 c = 10 b = 4|''
  \item ``\verb|a = 5 d = 10 c = 4 b = 6|''
  \item ``\verb|b = 6 d = 10 c = 4 a = 5|''
\end{enumerate}

\item \textbf{Task 10 of 16: The ``\{ \}'' symbol represents interpolation. Press 1, 2, 3 or 4 to select the correct interpolation.}\\
\begin{verbatim}
a = 5
b = 6
c = 4
d = 10

"a = {a} b = {d} c = {b}  d = {c}"
\end{verbatim}
\emph{Options:}
\begin{enumerate}
  \item ``\verb|a = 4 b = 5 c = 6  d = 10|''
  \item ``\verb|a = 5 b = 10 c = 6  d = 4|''
  \item ``\verb|a = {a} b = {d} c = {b}  d = {c}|''
  \item ``\verb|a = a b = b c = b  d = c|''
\end{enumerate}

\item \textbf{Task 11 of 16: The ``+'' symbol represents concatenation. Press 1, 2, 3 or 4 to select the correct concatenation.}\\
\begin{verbatim}
"(0 " + " 1)" + "(1 " + " 1)" +...+ m + n-1 + l-2 = m(m+1)/2 = O(m^2)
\end{verbatim}
\emph{Options:}
\begin{enumerate}
  \item ``\verb|(01)(11) ... mn-1l-2 = m(m1)/2 = O(m^2)|''
  \item ``\verb|12...mn-1l-2=mm1/2=0m^2|''
  \item ``\verb|(0 1) (1 1) ... m n-1 l-2 = m(m 1)/2 = O(m^2)|''
  \item ``\verb|1 2 ... mn-1l-2 = mm1/2 = 0m^2|''
\end{enumerate}

\item \textbf{Task 12 of 16: The ``\{ \}'' symbol represents interpolation. Press 1, 2, 3 or 4 to select the correct interpolation.}\\
\begin{verbatim}
"{(0 2 1)} {(1 2 1)} {...} n {m-1} {l-2} = m(m 1)/2 = O(m^2)"
\end{verbatim}
\emph{Options:}
\begin{enumerate}
  \item ``\verb|(0 2 1) (1 2 1) ... n m-1 l-2 = m(m 1)/2 = O(m^2)|''
  \item ``\verb|3 4 ... n m-1 l-2 = m(m 1)/2 = O(m^2)|''
  \item ``\verb|(021)(121) ... nm-1l-2 = m(m 1)/2 = O(m^2)|''
  \item ``\verb|... n m-1 l-2 = m(m 1)/2 = O(m^2)|''
\end{enumerate}

\item \textbf{Task 13 of 16: The ``+'' symbol represents concatenation. Press 1, 2, 3 or 4 to select the correct concatenation.}\\
\begin{verbatim}
a = "Hi"
b = "There"
userName = "Peter"

"C:\\Users\\" + userName + "\\Documents"
\end{verbatim}
\emph{Options:}
\begin{enumerate}
  \item ``\verb|Hi There Peter|''
  \item ``\verb|C:\\Users\\Documents\\Peter|''
  \item ``\verb|C:\\Peter\\Users\\Documents|''
  \item ``\verb|C:\\Users\\Peter\\Documents|''
\end{enumerate}

\item \textbf{Task 14 of 16: The ``\{ \}'' symbol represents interpolation. Press 1, 2, 3 or 4 to select the correct interpolation.}\\
\begin{verbatim}
a = "Hi"
doc = "Downloads"
userName = "Peter"

"C://Users//{doc}//{userName}"
\end{verbatim}
\emph{Options:}
\begin{enumerate}
  \item ``\verb|C://Users//Peter//Peter|''
  \item ``\verb|C://Users//Downloads//Peter|''
  \item ``\verb|C://Users//Peter//Downloads|''
  \item ``\verb|C:\\Users\\Peter\\Documents|''
\end{enumerate}

\item \textbf{Task 15 of 16: The ``+'' symbol represents concatenation. Press 1, 2, 3 or 4 to select the correct concatenation.}\\
\begin{verbatim}
domain = "example.com"
query = "?user=john"
protocol = "https://"
path = "/home"

protocol + domain + path + query
\end{verbatim}
\emph{Options:}
\begin{enumerate}
  \item ``\verb|https://example.com/home?user=john|''
  \item ``\verb|https://example.com /home ?user=john|''
  \item ``\verb|https://example.com/home|''
  \item ``\verb|https://example.com?user=john|''
\end{enumerate}

\item \textbf{Task 16 of 16: The ``\{ \}'' symbol represents interpolation. Press 1, 2, 3 or 4 to select the correct interpolation.}\\
\begin{verbatim}
domain = "example.com"
query = "?user=john"
protocol = "https://"
path = "/home"

"https://protocol/{domain}/path/{query}"
\end{verbatim}
\emph{Options:}
\begin{enumerate}
  \item ``\verb|https://protocol/home/?user=john|''
  \item ``\verb|https://example.com//home/?user=john|''
  \item ``\verb|https://protocol/example.com/path/?user=john|''
  \item ``\verb|https://https:///example.com//home/?user=john|''
\end{enumerate}

\end{enumerate}
\end{llmprompt}

\begin{llmprompt}[Open-ended Qualitative Survey Questions]{prompt:survey_qual_questions}
\footnotesize
\begin{enumerate}
    \item Based on your experience with both string interpolation and string concatenation in the survey, which approach did you find more readable, and why? Please provide specific examples or reasons to support your viewpoint.

    \item Did using string interpolation or concatenation affect your ability to quickly understand or debug code snippets? Please describe any challenges or advantages you experienced.

    \item If you were tasked with choosing between string interpolation and string concatenation for future programming tasks, which would you prefer, and what specific factors influenced your choice?

    \item How did your prior familiarity or experience with string interpolation or concatenation impact your ease of understanding the provided code examples? Please elaborate on your experience.

    \item Do you have any suggestions or recommendations on how either string interpolation or concatenation syntax could be improved to enhance readability and comprehension?
\end{enumerate}
\end{llmprompt}

\end{document}